\documentclass[lettersize,journal]{IEEEtran}
\usepackage{amsmath,amssymb,amsfonts}
\usepackage{algorithmic}
\usepackage[ruled]{algorithm2e}
\usepackage{array}
\usepackage{textcomp}
\usepackage{stfloats}
\usepackage{url}
\usepackage{color}
\usepackage{verbatim}
\usepackage{graphicx}
\usepackage{tabularx}
\usepackage{multirow}
\usepackage{array}
\usepackage{booktabs}
\usepackage{graphicx}
\usepackage{parskip}

\newtheorem{definition}{Definition}
\usepackage[numbers]{natbib}
\usepackage{bm}
\usepackage{mathtools}

\usepackage{subfigure}
\usepackage{makecell}

\usepackage{bibspacing}
\graphicspath{ {./fig/} }

\usepackage{flushend}

\def\BibTeX{{\rm B\kern-.05em{\sc i\kern-.025em b}\kern-.08em
    T\kern-.1667em\lower.7ex\hbox{E}\kern-.125emX}}
\usepackage{balance}
\begin{document}
\title{Networked Physical Computing: A New Paradigm for Effective Task Completion via Hypergraph Aided Trusted Task-Resource Matching}
\author{Botao~Zhu,~\IEEEmembership{Member,~IEEE} and Xianbin~Wang,~\IEEEmembership{Fellow,~IEEE}
\thanks{
B. Zhu and X. Wang are with the Department of Electrical and Computer Engineering, Western University, London, Canada N6A 5B9. Emails: \{bzhu88, xianbin.wang\}@uwo.ca.}
}


\maketitle

\begin{abstract}


Due to the diverse physical attributes of computing resources and tasks, developing effective mechanisms to facilitate task and resource matching in complex connected systems for value-oriented task completion has become increasingly challenging. To address the challenge, this paper proposes a networked physical computing system that integrates the physical attributes of computing resources and tasks as well as task-specific trust relationships among devices to enable value-driven task completion. Specifically, we propose a state-of-the-art hypergraph-aided trusted task-resource matching (TTR-matching) framework to achieve the envisioned physical computing. First, a task-specific trusted physical resource hypergraph is defined, which integrates task-specific trust, the physical attributes of resources, and task types. This enables accurate modeling of device collaboration dependencies under specific task types. Next, a task hypergraph is generated to associate the task initiator with the physical attributes of the corresponding tasks. Based on these two hypergraphs, a hypergraph matching algorithm is designed to facilitate task-specific trusted collaborator selection and accurate task-resource matching for value-maximizing task completion. Extensive experimental results demonstrate that the proposed TTR-matching framework outperforms comparison algorithms in identifying task-specific trustworthy collaborators and maximizing the average value of task completion.







\end{abstract}

\begin{IEEEkeywords}

Networked physical computing, hypergraph, task-specific trust, task-resource matching, value of task completion. 

\end{IEEEkeywords}

\section{Introduction}

\IEEEPARstart{D}{ue} to the growing complexity of applications and connected systems, it becomes impractical for each individual device to handle all related computing tasks by itself using its limited computation capacity and energy supply~\cite{9831429},~\cite{9906430}. To address this challenge, many researchers have started exploring the collaborative use of distributed resources of peer devices to enable cost-effective task execution through distributed computing~\cite{8632974},~\cite{10453339}. Currently, a wide range of applications based on device collaboration is emerging across various industry sectors, aimed at achieving low-latency and high-quality task completion. For instance, autonomous driving leverages collaborative systems to ensure safe and efficient navigation through coordinated vehicle-to-vehicle communication~\cite{10177684}. Similarly, in the metaverse, device collaboration plays a central role in delivering immersive, seamless, and high-performance virtual experiences, supporting high-resolution rendering, real-time interaction, and responsive user interfaces~\cite{10144502}. These systems enable seamless interaction and efficient resource use, shaping future interconnected environments. 



\subsection{Motivations}

The goal of collaborative computing is to ensure effective task completion by leveraging distributed resources. Achieving this goal relies on  the selection of appropriate collaborators according to tasks requirements. Given the complexity and heterogeneity of the physical attributes of both tasks and resources, effective collaborator selection mechanisms are essential to ensure reliable and efficient task completion.

Recent research has explored trust as a key metric for collaborator selection by reliability evaluation. In this context, trust is defined as the task owner's expectation of a collaborator’s capability, inferred from factors such as historical behavior and available resources~\cite{icc_botao_2},~\cite{10320384},~\cite{10104094}. However, such generalized trust assessments fall short in collaborative environments involving heterogeneous tasks and distributed resources. In these scenarios, trust must be evaluated in a task-specific manner to capture how appropriate a collaborator can meet the unique requirements of each individual task~\cite{chain_of_trsut}. Therefore, developing task-specific trust evaluation mechanisms is essential for accurately identifying trustworthy collaborators.

Furthermore, the physical attributes of devices, networks, and tasks are inherently heterogeneous and diverse within complex systems, which have to be considered in task-resource matching for task execution outcome. The physical attributes of devices, such as CPU, storage, memory, and energy consumption, directly determine their ability to execute tasks. Due to differences in physical performance, different devices may exhibit varying efficiency and effectiveness when processing tasks, which increases the complexity of resource allocation~\cite{8516361},~\cite{8933071}. In addition, network attributes, including transmission speed, jitter, and topology, play a critical role in influencing task distribution and system-wide scheduling decisions~\cite{8340067},~\cite{9999298}. For example, high transmission speed and well-structured network topology can improve transmission reliability and reduce latency, thereby facilitating more efficient task execution across the system. Furthermore, as tasks become more complex, they impose diverse demands on the capabilities of collaborators~\cite{9275320},~\cite{10623874}, such as computational power, network performance, energy, and security. These task requirements have a direct impact on how resources are distributed and significantly impact the overall effectiveness of task execution. However, the physical attributes of resources and tasks have been largely overlooked in prior research, leaving a critical gap in understanding how their differences shape task outcomes in collaborative systems. Therefore, task-resource matching mechanisms that account for the heterogeneity in physical attributes of both resources and tasks are urgently needed to ensure effective task completion.

Based on the above analysis, achieving effective task completion in collaborative systems depends not only on selecting the right collaborators for each task but also on accurately matching tasks with appropriate resources. Yet, existing research has paid limited attention to mechanisms that jointly address these two dimensions. To bridge this gap, this work introduces a networked physical computing system that incorporates the physical attributes of both tasks and devices into the matching process. It further builds task-specific trusted collaboration relationships among devices, enabling more accurate collaborator selection and value-driven task completion.

\subsection{Key Technical Challenges}

For the successful deployment and operation of the networked physical computing system, the following technical challenges need to be addressed. First of all, how trust can be used as a tool to enhance the allocation of resources and tasks requires careful consideration. In the networked physical computing system, trust between any pair of devices is not a single value, but a dynamic and diverse relationship. This dynamic nature and diversity mainly arise from the heterogeneity of devices \cite{9262001}. For instance, a device that uses the same physical resources to perform different tasks may exhibit varying performance for each task, leading to diverse levels of trust~\cite{icc_botao}. Therefore, trust between devices should not only be task-specific but also evolve over time. Accurately capturing these task-specific trust relationships and leveraging them to guide the allocation of resources and tasks are essential for the operation of the networked physical computing system.

In addition, the precise matching of resources and tasks presents another critical challenge. Apart from the heterogeneity and complexity of device resources, task attributes are also complex \cite{9760005}, i.e., different tasks impose varying requirements on resources. For example, computation-sensitive tasks—such as image recognition, video processing, and fault prediction—typically require high processing power and low latency. However, user devices are often resource-constrained and incapable of handling such tasks independently. Therefore, these tasks must be offloaded to devices with sufficient computational capabilities. Effective task–resource matching requires in-depth comparative analysis of both the physical characteristics of tasks—such as data volume, computational complexity, and real-time requirements—and the physical attributes of devices, including CPU/GPU performance, memory capacity, network latency, and current load. Arbitrary assignment could lead to execution failures, delayed responses, and system congestion. Hence, designing effective mechanisms to facilitate task-resource matching—while taking into account the physical attributes of both tasks and resources—is essential for improving the efficiency and reliability of task execution in complex computing environments.


Moreover, the effective coordination of physical resources, tasks, and trust is essential for task completion. Current research often treats these elements as independent or fails to explore their interrelationships. In fact, there are collaborative dependencies between devices under specific tasks, which we refer to as high-order relationships. Specifically, a high-order relationship between any pair of devices involves two devices, a specific task, and the trust between them. Unlike the traditional generalized definitions of device cooperation relationships, the high-order relationships can provide a more detailed, precise, and comprehensive characterization of inter-device collaboration, offering stronger support and assurance for task and resource allocation. Therefore, accurately uncovering the high-order relationships between resources, tasks, and trust is crucial for the development of the networked physical computing system.

Furthermore, to implement the proposed networked physical computing system, it is essential to identify an effective tool that can seamlessly integrate resources, tasks, task-specific trust relationships, and the task-resource matching mechanism into a cohesive framework. This integration will not only streamline the coordination of these components but also enhance the system’s overall performance.

Recent studies have demonstrated that hypergraphs are a powerful tool for effectively capturing multiple relationships in complex networks~\cite{10546264}. Hypergraphs offer a more expressive and flexible representation by permitting multiple nodes to be connected by a single hyperedge. This enhanced flexibility allows for a more comprehensive depiction of complex relationships among different elements. Therefore, hypergraphs can be used to represent combinations of device resources, tasks, and trust in the networked physical computing system, providing a thorough perspective on how devices interact and collaborate to accomplish tasks. In addition, due to the fact that hypergraphs can facilitate the representation of high-order relationships, they can play a pivotal role in collaborative decision-making processes, particularly in optimizing resource allocation and achieving effective task completion.

\subsection{Contributions}
Considering the challenges outlined in the networked physical computing system and the advantages of hypergraphs, this study proposes a hypergraph-aided trusted task-resource matching (TTR-matching) mechanism. The main contributions are summarized as follows:
\begin{itemize}
  \item
   This paper proposes an innovative networked physical computing system that leverages task-specific trusted relationships among devices to achieve precise task-resource matching based on their physical attributes.

   \item
   A task-specific trust model is proposed to effectively capture the dynamic and diverse nature of trust between devices. It provides a deeper understanding of how trust varies based on specific tasks, enabling the accurate representation of trust relationships among devices. 

   \item 
   We innovatively utilize hypergraphs to integrate task-specific trust, the physical attributes of resources and tasks. This integration enables a clearer and more accurate depiction of device collaboration dependencies under specific tasks, thereby promoting more effective cooperation.

   \item 
To ensure precise task-resource matching, a hypergraph-aided matching method is proposed to align tasks and resources between the task hypergraph and the task-specific trusted physical resource hypergraph to maximize the average value of task completion.

 \item
 Experimental results demonstrate that the proposed TTR-matching model effectively captures task-specific trust relationships among devices and outperforms other methods in maximizing the average value of task completion.


    


\end{itemize}

The rest of the paper is organized as follows. A comprehensive review of the literature is provided in Section \ref{related_work}. The networked physical computing system model is introduced in Section \ref{Modeling of Trusted Collaborative IoT system}. Task-specific trusted physical resource hypergraph is presented in Section \ref{hypergraph-task-specific}, and resource-task matching is proposed in Section \ref{section_hypergraph_matching}. Simulation results are provided in Section \ref{simulation_discussion}. Finally, Section \ref{conclusion} concludes this paper.

\section{Related Work}
\label{related_work}
In this section, we will first conduct an in-depth review of existing research on resource optimization in collaborative systems with heterogeneous resources and tasks, followed by a summary of their shortcomings. Next, we will investigate trust-enabled collaborative systems and analyze their limitations. A comparison of our work with existing collaborative systems across multiple dimensions is provided in Table~\ref{related_comparison}. 

\subsection{Resource Optimization in Collaborative Systems with Heterogeneous Resources and Tasks}

In recent years, the impact of the heterogeneous and complex attributes of resources and tasks on resource optimization in collaborative systems has garnered attention. Some researchers have integrated computing, communication, and network resource attributes into the development of resource optimization and task offloading strategies. In~\cite{9749874}, a distributed orchestration mechanism was proposed to achieve reliable collaborative rendering in a dynamic Internet of Things (IoT) system by recognizing the actual resource condition and understanding the mutual impact of resource condition and task performance. In~\cite{10024361}, the authors presented a collaborative cloud-edge computation offloading model for a heterogeneous environment with multiple users, multiple edge servers and multidimensional resource requirements. In~\cite{10623874}, the authors investigated the offloading strategy in the edge computing task scenario by considering the characteristics of terminal devices, workflow task attributes, and network resources. In~\cite{9672105}, the authors studied the joint communication and computation resource allocation for computation offloading in fog-based vehicular networks. To reduce the communication cost, a model update compression scheme was introduced to enable communication-efficient decentralized federated learning for artificial intelligence of things in~\cite{9785702}.

\vspace{-0.015 in}
Additionally, some researchers have designed task-oriented offloading strategies in collaborative systems by considering task attributes and requirements. In~\cite{9931975}, the authors developed a task offloading approach within a three-layer end-edge-cloud framework by taking task attributes into account, such as input data size, total CPU cycles, and maximum completion deadline. In~\cite{9783171}, the authors used a direct acyclic graph to model the relationships between tasks and design the optimal offloading sequence for a set of tasks in an edge computing system. In~\cite{9967961}, considering data size, computation workload, and delay requirement as the main attributes of tasks, the authors proposed a task offloading strategy that maximizes platform revenue in a device-to-device cooperative system. In~\cite{9275320}, by considering task attributes, such as the input size and the output size of the task, computing resource requirement, time constraint, and the popularity of task result, the authors proposed a two-stage task offloading mechanism to provide computing services in a collaborative vehicular network. 
The authors in \cite{9810544} classified tasks within the Internet of Vehicles network into common and urgent categories and proposed a task offloading approach aimed at minimizing energy consumption by considering the varying requirements of different tasks.

However, the aforementioned studies exhibit two critical shortcomings. First, they neglect the alignment between tasks and resources, potentially causing inefficient utilization of available resources. Second, they fail to incorporate trust evaluation in the collaborator selection process, which risks task execution failure if untrustworthy collaborators are chosen.

\subsection{Trust-Enabled Collaborative Systems}

In recent years, researchers have explored diverse trust factors and proposed various trust evaluation approaches to effectively identify trustworthy collaborators in collaborative systems. In~\cite{10422726}, the authors proposed a dynamic trust management model based on feedback for edge devices in the industrial internet. In~\cite{10103199}, they calculated trust based on task completion feedback and used it to select trustworthy collaborators in the Internet of Vehicles (IoVs) network. In~\cite{10320384}, the authors implemented a rapid and reliable trust evaluation method based on historical experience and real-time resource conditions to select collaborators in the IoVs. The authors in \cite{10251781} evaluated the trust levels of edge computing servers according to their bandwidth, computing ability, and initial success rate, and offloaded tasks to trusted servers. In~\cite{9794601}, a semi-centralized trust management system was developed, which calculates trust based on direct experiences and indirect recommendations. In~\cite{9305298}, device trust was evaluated by integrating computation capabilities, social relationships, external opinions, and dynamic knowledge. To select reliable collaborators, the authors emphasized node reputation as a critical factor in assessing trustworthiness in~\cite{9908525}.

In recent years, some researchers have investigated the application of trust in collaborative computing systems to ensure successful task completion. In \cite{10158489}, the authors proposed a two-tier trust evaluation model that not only evaluates the trust of underlying data reporters, but also detects malicious attackers in the routing layer in an edge-cloud computing system, thereby increasing malicious node detection rate. In~\cite{10104094}, to process virtual reality video streaming tasks collaboratively, the authors presented a novel trust evaluation method by combining direct and indirect trust values, and jointly optimized trusted collaborator selection, spectrum allocation, and rendering resource allocation in a mobile edge computing system. In \cite{10843332},  the authors evaluated mobile users' trust based on direct interactions and recommendations. They then designed a trust-aware task offloading and resource allocation mechanism to reduce task scheduling latency while maintaining acceptable system performance in mobile cloud-edge computing networks.

However, the above studies on trust have the following limitations. First, they primarily focus on trust evaluation based on factors such as feedback, historical interactions, and social relationships, without establishing trusted cooperative relationships among devices. Second, they consider only one-to-one trust between devices and overlook the task-specific diversity of trust.

\subsection{Hypergraph Application in Wireless Systems}
Hypergraphs are a special type of graph capable of representing more intricate relationships among multiple entities than conventional graphs. This enhanced expressiveness makes them particularly suitable for addressing problems in complex systems. In recent years, researchers have begun exploring the application of hypergraphs in wireless systems, aiming to leverage their high-order relational modeling capabilities to enhance system performance. To mitigate intra-cluster pilot contamination caused by non-orthogonal multiple access (NOMA), which significantly degrades system performance, a novel hypergraph coloring-based algorithm was proposed in \cite{10972056} for pilot assignment by modeling the cumulative interference among multiple users. In this algorithm, the interference hypergraph was constructed using a main access point selection strategy, and pilots were allocated based on the interference weights within the hypergraph. In \cite{8491251}, the authors studied spectrum allocation using hypergraph coloring method in uplink device-to-device (D2D) underlaying cellular networks to eliminate cumulative interference from multiple D2D pairs and enhance the system capacity. To maximize the system sum-rate in a RIS-assisted D2D multicast system, the authors in \cite{10444009} first developed an efficient channel allocation algorithm using hypergraph and then jointly optimized the channel reuse coefficients, active beamforming at the access point, and passive beamforming at the RIS. In \cite{9990528}, the authors proposed a new hypergraph-based network slicing framework that utilizes hypergraph theory to process numerous interlinked logical networks and interconnected physical infrastructures. Inspired by the success of these studies, this paper further investigates the application of hypergraphs to improve wireless systems, with a focus on resource allocation in collaborative environments.

\begin{table}[!t] 
        \footnotesize  
	\centering
        \renewcommand{\arraystretch}{1.3}
	\caption{A comparison of six dimensions between our work and existing collaborative systems.}
	\label{related_comparison}
	  \begin{tabular}{p{0.7cm}<{\centering}|p{0.8cm}<{\centering}|p{0.8cm}<{\centering}|p{0.8cm}<{\centering}|p{0.8cm}<{\centering}|p{0.8cm}<{\centering}|p{0.7 cm}<{\centering}}
		\hline {Ref.} & {DA} & {TA} & {NA} & {TC} & {TST}  & {TRM} \\
		\hline \hline
              \cite{9749874}  & $\surd$  & $\surd$  & $\surd$ & $\times$ &  $\times$   & $\times$ \\
             \hline
             \cite{10024361}  &  $\surd$  &  $\surd$  &  $\surd$ & $\times$ &  $\times$    &  $\times$  \\
             \hline
            \cite{10623874}   &  $\surd$  &  $\surd$  &  $\surd$ & $\times$ &  $\times$   & $\times$   \\
             \hline
            \cite{9672105}  &   $\surd$   &  $\surd$  &  $\surd$ &  $\times$ &  $\times$    & $\times$  \\
             \hline
            \cite{9785702}  &  $\times$  &  $\times$  &  $\surd$ & $\times$ &  $\times$      & $\times$     \\
             \hline
            \cite{9931975}  & $\surd$    &  $\surd$   & $\surd$  & $\times$ &  $\times$    &  $\times$ \\
             \hline
            \cite{9783171}  &  $\times$  &  $\surd$  & $\times$ &  $\times$ &  $\times$   &  $\times$  \\
             \hline
            \cite{9967961} &  $\surd$   &  $\surd$   & $\surd$  & $\times$ &  $\times$    & $\times$   \\
             \hline
            \cite{9275320}  &  $\surd$  &  $\surd$  & $\surd$ & $\times$ & $\times$     &  $\times$ \\
             \hline
            \cite{9810544}  &  $\surd$  &  $\surd$  &  $\surd$ & $\times$ &  $\times$   & $\times$  \\
             \hline
            \cite{10422726}  &   $\times$ &  $\times$  & $\times$ & $\surd$ &  $\times$    &  $\times$  \\
             \hline
            \cite{10103199}  & $\surd$   &  $\times$  & $\surd$ & $\surd$ & $\times$     &  $\times$ \\
             \hline
             \cite{10320384}  &  $\surd$  &  $\surd$  & $\surd$ & $\surd$ &  $\times$    & $\times$  \\
             \hline
             \cite{10104094} &  $\surd$   &  $\surd$   & $\surd$  & $\surd$  &  $\times$   & $\times$   \\
             \hline
             \cite{10251781}  &   $\surd$  &   $\surd$  &  $\surd$ &  $\surd$ &  $\times$  & $\times$  \\
             \hline
            \cite{9794601}   &   $\times$  &  $\times$   &   $\times$ &  $\surd$ &   $\times$    &  $\times$   \\
             \hline
            \cite{9305298}   &  $\surd$  &  $\times$  &  $\times$  &  $\surd$ &   $\times$   & $\times$  \\
             \hline
             \cite{9908525}   &  $\times$  &  $\times$  &  $\times$  &  $\surd$ &   $\times$   & $\times$  \\
             \hline
             \textbf{Our work}  & $\surd$ &  $\surd$ & $\surd$ & $\surd$ & $\surd$ & $\surd$   \\
             \hline
             \multicolumn{7}{l}{Note: $\surd$ indicates that this factor is considered in collaborative systems,}\\
             \multicolumn{7}{l}{ while $\times$ indicates that it is not. DA: device attribute, TA: task attribute,}\\
             \multicolumn{7}{l}{ NA: network attribute, TC: trusted collaboration, TST: task-specific}\\
             \multicolumn{7}{l}{ trust, TRM: task-resource matching.}\\
             \hline
   \end{tabular}
\end{table}

\section{Networked Physical Computing System Model}
\label{Modeling of Trusted Collaborative IoT system}

\begin{figure}[t!]
\centering
\includegraphics[scale=0.78]{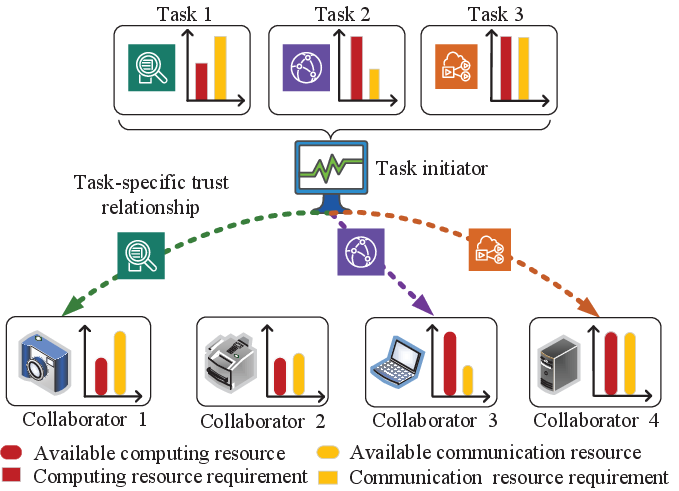}
\caption{Task-resource matching based on task-specific trust and physical attributes of resources and tasks in a networked physical computing system.}
\label{system model}
\end{figure}

The core of the networked physical computing system lies in leveraging trust to establish task-specific cooperative relationships among devices. By integrating these relationships with the physical attributes of tasks and resources, the system facilitates more effective and accurate task–resource matching. As illustrated in Fig.~\ref{system model}, different tasks have varying demands for computing and communication resources, while different devices offer diverse capabilities. The established task-specific trust relationships among devices enable the system to effectively align task requirements with the available resources, ensuring optimal allocation and performance.
To implement this system, we first introduce the physical resource and task models. Then, we discuss the definition of trust in the proposed system and explore how trust can be used to establish task-specific cooperative relationships among devices by proposing a task-specific trust model. In addition, the collaboration model among devices is introduced. Furthermore, the value of task completion is defined to measure the effectiveness of collaboration. Finally, we formulate a task-resource matching problem aimed at maximizing the value of task completion. The main notations used in this research are summarized in Table~\ref{notation}.

\subsection{Physical Resource and Task Models}

 All devices in the system are represented as $\bm{A} = \{a_1, \dots, a_J\}$. For simplicity, each device $a_j, j=1,\dots, J$, is assumed to be equipped with a single-core CPU and a single antenna. Device $a_j$ is parameterized as $(id_{a_j}, f_{a_j}, g_{a_j}, p_{a_j}, \bm{S}_{a_j})$, where $id_{a_j}$ is the unique identifier of device $a_j$,  $f_{a_j}$ represents the CPU clock frequency, $g_{a_j}$ is the coordinate of device $a_j$ in physical space, $p_{a_j}$ is the transmission power, and $\bm{S}_{a_j}$ denotes the set of task types supported by device $a_j$. A task type refers to a specific application, and a device supporting a task type means that it can utilize its software and hardware resources to execute tasks of this type. For example, facial recognition can be considered a task type, and a device supporting this task should be equipped with a high CPU frequency and the necessary facial recognition software. The set of all task types in the network is defined as $\bm{S} = \bm{S}_{a_1}\cup \dots \cup \bm{S}_{a_J} = \{1,\dots, S\}$. It should be noted that in the assumed device model, different devices exhibit varying capabilities when processing the same task, which aligns with the heterogeneity observed in real-world systems. On one hand, devices have different CPU frequencies, resulting in different execution times and energy consumption for the same task. On the other hand, the types of tasks a device can support vary depending on its installed software and hardware resources. Therefore, a given task may be executable on some devices but not on others.

  Any device $a_i \in \bm{A}$ can act as a task initiator to generate a task $\bm{B}$, which consists of a series of subtasks $\{b_1,b_2,\dots,b_M\}$ that can be executed independently. 
  To simplify the complexity, each subtask $b_m, m=1,\dots, M$, $M < J$, is characterized as $(s,\rho_{b_{m}}, d_{b_m}, t^{\text{max}}_{b_m}, T_{b_m}, r_{b_m})$, where $s$ refers to the task type of $b_{m}$, $\rho_{b_m}$ represents the processing density, defined as the number of CPU cycles needed to process a unit bit (cycles/bit)~\cite{10546264}, $d_{b_m}$ is the number of data bits, $t_{b_m}^{\text{max}}$ is the maximum task completion tolerance time, $T_{b_m}$ denotes the minimum trust demand for collaborators, and $r_{b_m}$ is the minimum transmission rate demand for communication links.

  Here, we clarify the distinction between the concepts of task and task type.  A task type refers to a particular application that is determined by the software and hardware resources of a device, whereas a task encompasses the specific requirements associated with that application, which is determined by the task initiator~\cite{10594714}. For example, we assume that subtask $b_m$ is a facial recognition task, then its task type is facial recognition and its specific task information can include a maximum task completion tolerance time of 1 second, a task size of 20 MB, and a minimum trust demand of 0.5.  If subtask $b_m$ is assigned to device $a_j$, the resources of device $a_j$ can support the execution of tasks of facial recognition type. For tasks of the same type, their requirements may vary.

\begin{table}[!t] 
	\centering
        \renewcommand{\arraystretch}{1.2}
	\caption{Summary of Notation}
	\label{notation}
	  \begin{tabular}{p{0.9cm}<{}|p{7.1cm}<{}}
		\hline \textbf{Notation} & \textbf{Description} \\
		\hline \hline
              $\bm{A}$ & The set of devices\\  
              \hline
		   $\bm{B}$ & Task \\ 
              \hline
              $b_m$ & One subtask of $\bm{B}$\\
              \hline
              $s$ & Task type \\
              \hline
              $T_{a_j}$ &  The overall trust that all relevant devices in the system have in device $a_j$ \\
              \hline
              $T_{a_i, a_j}$ &  The direct trust of device $a_i$ to device $a_j$\\
              \hline
              $T_{a_i,s,a_j}$ & The direct trust of device $a_i$ to device $a_j$ specific to task type $s$
              \\
              \hline
              $T^s_{a_i,a_j}$ & The trust of device $a_i$ to device $a_j$ specific to task type $s$\\
              \hline
              $t_{b_m}$ & The total task time of  subtask $b_m$ \\
              \hline
              $E_{b_m}$ & The total energy consumption for completing subtask $b_m$\\
              \hline
              $V_{b_m}$ & Value of task completion\\
              \hline
              $\bm{C}_s$ &  The cluster of devices that support task type $s$ \\
              \hline
              $\mathcal{H}^{\text{grp}}$ &  Group trust hypergraph\\
              \hline
              $e^s_{a_j}$ &  A hyperedge in $\mathcal{H}^{\text{grp}}$ where device $a_j$ is the central node \\
              \hline
              $\mathcal{G}^{\text{dec}}$ & Directed graph \\
              \hline
              $e^{\text{dec}}_{a_i \to a_j}$ & The directed edge from device $a_i$ to device $a_j$ \\
              \hline
              $w_{a_i \to a_j}$ & The weight of $e^{\text{dec}}_{a_i \to a_j}$ \\
              \hline
              $\mathcal{H}^{\text{res}}$ &  Task-specific trusted physical resource hypergraph\\
              \hline
              $e^s_{a_i \to a_j}$ & The hyperedge connecting device $a_i$, device $a_j$, and task type $s$ \\
              \hline
              $w^s_{a_i \to a_j}$ & The weight of $e^s_{a_i \to a_j}$ \\
              \hline
              $\mathcal{H}^{\text{task}}$ & Task hypergraph\\
              \hline
              $e^{s'}_{a_i' \to \phi}$ & The task hyperedge connecting device $a'_i$ and subtask $b_m$ with task type $s'$\\
              \hline
              $w^{s'}_{a_i' \to \phi}$ & The weight of $e^{s'}_{a_i' \to \phi}$ \\
              \hline
              $\mathcal{E}^{\text{task}}$ & The set of hyperedges of $\mathcal{H}^{\text{task}}$ \\
              \hline
              $\mathcal{E}^{\text{res}}$ & The set of hyperedges of $\mathcal{H}^{\text{res}}$ \\
              \hline
              $\bm{N}$ & The set of strategies, and each strategy is the candidate match between a task hyperedge and a resource hyperedge \\
            \hline
   \end{tabular}
\end{table}
  
\subsection{Definition of Trust and Task-Specific Trust Model}

Device $a_i$, as the task initiator, needs to find a suitable collaborator for each subtask, which requires assessing the trustworthiness of potential collaborators. Collaborators with low trust levels are more likely to lead to task failure. Therefore, accurately assessing their trustworthiness is essential. We first define trust as follows:
\begin{definition}[Trust in the networked physical computing system] 
\textit{The trust of the task initiator $a_i$ towards a potential collaborator $a_j$ is expressed as $a_i$'s expectation of $a_j$ to complete a subtask, which is assessed based on $a_j$'s communication, computing, and other capabilities.} 
\end{definition}

\begin{figure}[t!]
	\centering
	\includegraphics[scale=0.8]{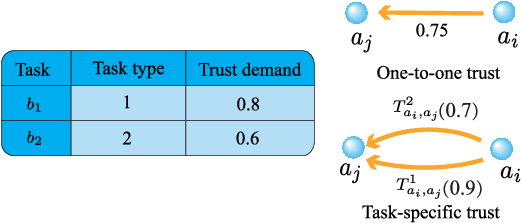}
	\caption{Comparison of the one-to-one trust model and the task-specific trust model in achieving task collaboration.}
	\label{multi-tier trust}
\end{figure}

Based on the definition of trust, devices can establish mutual trust relationships with each other. However, existing research on trust generally assumes that there is only a one-to-one trust relationship between any pair of devices, which fails to truly reflect the actual trust relationship between them. A device should earn a high trust value when handling tasks it excels at, while it may receive a lower trust value when dealing with tasks it is less skilled at. Therefore, trust relationships between devices should be diverse and task-specific. This diversity and task-specific nature of trust will influence the selection of collaborators. Take Fig.~\ref{multi-tier trust} as an example. Device $a_i$ has two subtasks with different task types (type 1 and type 2), and these two subtasks have different trust demands for collaborators. If there is a one-to-one trust relationship from device $a_i$ to device $a_j$ with a value of 0.75, device $a_j$ can be selected as a collaborator to perform only subtask $b_2$, as this trust value is greater than the trust demand of $b_2$.
However, if device $a_i$ has two trust relationships with device $a_j$ specific to task types 1 and 2, where $T^1_{a_i,a_j}$ has a value of 0.9 and $T^2_{a_i,a_j}$ has a value of 0.7, then device $a_j$ is qualified to perform subtask $b_1$ based on $T^1_{a_i,a_j}$ and subtask $b_2$ based on $T^2_{a_i,a_j}$.



It can be observed that task-specific trust effectively facilitates collaboration among devices by aligning task requirements with the capabilities of collaborators. Therefore, in the networked physical computing system, task-specific trust is utilized to establish reliable cooperative relationships between devices. Formally, the trust of device $a_i$ to device $a_j$ specific to task type $s$ is defined as
\begin{align}
    \label{task-specific}
    T_{a_i,a_j}^{s} = \beta_1 T_{a_j} + \beta_2 T_{a_i,a_j} + \beta_3 T_{a_i,s,a_j},
\end{align}
where $\beta_1$, $\beta_2$, and $\beta_3$ are the weight parameters, and $\beta_1 + \beta_2 + \beta_3 = 1$, $0 \le \beta_1, \beta_2, \beta_3 \le 1$. $T_{a_j}$ represents the overall trust that all relevant devices in the system have in device $a_j$, $T_{a_i, a_j}$ represents the direct trust of device $a_i$ to device $a_j$ via evaluating their historical interactions, interests, and other factors. 
$T_{a_i,s,a_j}$ represents the direct trust of device $a_i$ in device $a_j$ with respect to task type $s$, derived from evaluating device $a_j$’s performance in executing tasks of type $s$ assigned by device $a_i$. Based on task-specific trust, device $a_i$ can accurately select a trusted collaborator for each subtask. 

According to the task-specific trust model defined in equation~(\ref{task-specific}), the trust between any pair of devices exhibits both diversity and dynamism. Regarding trust diversity, when device $a_j$ performs multiple tasks of different types assigned by device $a_i$, device $a_i$ evaluates the trust of device $a_j$ for each task type separately, thereby establishing multiple task-specific trust relationships between them. As for trust dynamism, each time device $a_j$ completes a task assigned by device $a_i$, device $a_i$ updates its trust in device $a_j$ based on the latest task execution result. This continuous evolution mechanism enables the trust evaluation to dynamically and promptly reflect the performance of devices.

\subsection{Collaboration Model}

Since task $\bm{B}$ of device $a_i$ consists of $M$ subtasks, it needs to select a suitable collaborator for each subtask. If device $a_j$ is selected as a potential collaborator to execute subtask $b_m$ with task type $s$, it is required to satisfy the following conditions: i) Device $a_j$ supports task type $s$, meaning that its hardware and software resources are sufficient to perform tasks of this type; ii) Its trust value meets or exceeds the minimum trust threshold $T_{b_m}$ required by subtask $b_m$; iii) The transmission rate of the link from device $a_i$ to device $a_j$ satisfies the minimum communication requirement $r_{b_m}$. Each subtask is offloaded to the selected collaborator and executed independently. There are no dependencies among subtasks. The collaboration process is divided into the task transmission phase and the task execution phase. During the task transmission phase, the task transmission time of $b_m$ from device $a_i$ to device $a_j$ is calculated by
\begin{align}
    t_{b_m}^{\text{tra}} = \frac{d_{b_{m}}}{r_{a_i, a_j}},
\end{align}
where $r_{a_i,a_j}$ is the transmission rate from device $a_i$ to device $a_j$. It is calculated by
\begin{align}
    r_{a_i,a_j} = W^{\text{band}} \log_2 \left(1 + \frac{p_{a_i}h_{a_i,a_j}}{N_0} \right),     
\end{align}
where $h_{a_i,a_j}$ is the channel gain, $N_0$ is the noise power, and $p_{a_i}$ is the transmission power of device $a_i$. A simple channel model $h_{a_i,a_j} = |g_{a_i}-g_{a_j}|^{-\alpha}$ is considered, where $|g_{a_i}-g_{a_j}|$ represents the distance between $a_i$ and $a_j$, and $\alpha=4$ denotes the path loss factor. 
The energy consumed in the transmission phase is calculated as
\begin{align}
    E^{\text{tra}}_{b_m} = p_{a_i}t^{\text{tra}}_{b_m}.
\end{align}
Device $a_j$ starts executing the subtask $b_m$ upon receiving it, and the task execution time is calculated as \cite{10546264}
\begin{align}
   t_{b_m}^{\text{exe}} = \frac{d_{b_{m}}\rho_{b_{m}}}{f_{a_{j}}}.
\end{align}
The energy consumed by device $a_j$ in executing subtask $b_m$ is given by
\begin{align}
\label{exe_energy}
    E^{\text{exe}}_{b_m} = \epsilon f^2_{a_j} d_{b_{m}}\rho_{b_{m}},
\end{align}
where $\epsilon f^2_{a_j}$ is the coefficient denoting the consumed energy per CPU cycle, and $\epsilon$ is set to $10^{-11}$ according to the measurements in \cite{6195685}. After completing subtask $b_m$, device $a_j$ returns the result to device $a_i$. It should be noted that the time and energy spent in establishing links and transmitting results between device $a_i$ and collaborators are omitted.

\subsection{Value of Task Completion as a Metric}

 To evaluate the performance of device $a_j$ in completing subtask $b_m$, a novel metric called the value of task completion is introduced. The concept of value has been widely employed in modelling market activities, representing a user's comprehensive assessment of the value derived from a product or service, considering their perception. Since users generally lack awareness of actual production or service costs, they make a purchase only when their perceived value for the product exceeds its selling price. 
 
\textit{1) Value of task completion}:
In this research, we define the value of task completion perceived by device $a_i$, the task initiator, as
\begin{align}
    \label{value}
    V_{b_m} = \xi_1 V^{\text{time}}_{a_i, b_m, a_j} + \xi_2 V^{\text{ener}}_{a_i, b_m, a_j},
\end{align}
where $\xi_1$ and $\xi_2$ are the weight parameters, $0 \le  \xi_1,\xi_2 \le 1$, $\xi_1 + \xi_2 = 1$. 
$V^{\text{time}}_{a_i,b_m,a_j}$ and $V^{\text{ener}}_{a_i,b_m,a_j}$ are the value of time and the value of energy, respectively. 

\textit{2) Value of time}:
Based on Kano's quality theory representing user satisfaction \cite{7852434}, $V^{\text{time}}_{a_i,b_m,a_j}$ is given by
\begin{align}
    V^{\text{time}}_{a_i,b_m,a_j} = \begin{cases}
    1, & \text{if} \  t^{\text{max}}_{b_m} \ge t_{b_m}; \\
    e^{-|{(t_{b_m} - t^{\text{max}}_{b_m})}/{t^{\text{max}}_{b_m}}|}, & \text{if} \ t^{\text{max}}_{b_m} < t_{b_m}, \\
   \end{cases}
\end{align}
where $t^{\text{max}}_{b_m}$ is the maximum task completion tolerance time, and $t_{b_{m}}$ is the completion time of subtask $b_{m}$, which is the sum of the task transmission time and the task execution time. If $t_{b_{m}}$ is greater than $t_{b_m}^{\text{max}}$, then $V^{\text{time}}_{a_i,b_m,a_j}$ is less than 1 and decreases as $t_{b_m}$ increases. 

\textit{3) Value of energy}: $V^{\text{ener}}_{a_i,b_m,a_j}$ is defined as the comparison between the expected energy consumption and the actual energy consumption, which is expressed as 
\begin{align}
    V^{\text{ener}}_{a_i,b_m,a_j} = \begin{cases}
    1, & \text{if} \ E^{\text{exp}}_{b_m} \ge E_{b_m}; \\
    e^{-|{(E_{b_m} - E^{\text{exp}}_{b_m})}/{E^{\text{exp}}_{b_m}}|}, & \text{if} \ E^{\text{exp}}_{b_m} < E_{b_m}, \\
   \end{cases}
\end{align}
where $E^{\text{exp}}_{b_m}$ is the expected energy consumption obtained by assuming subtask $b_m$ is completed by the task initiator $a_i$, and $E_{b_m}$ is the actual energy consumption when subtask $b_m$ is completed by device $a_j$. $E_{b_m}$ is the sum of the transmission energy consumption and the computation energy consumption.

\subsection{Problem Formulation of Task-Resource Matching}
According to (\ref{value}), the value of task completion gained from completing $\bm{B} = \{b_1, \dots, b_M \}$ is determined by the selected collaborators. The better the alignment between task requirements and the collaborator’s resources, the greater the value gained by the task initiator.
The allocation relationship between each subtask $b_m$ and each collaborator $a_j$ is assumed as $l_{a_j,b_m} = 1$ if $b_m$ is allocated to $a_j$, otherwise $l_{a_j,b_m} = 0$. To maximize the average value of task completion, the optimization problem is formulated as the precise matching of tasks and resources in the networked physical computing system 
\begin{alignat}{1}
    P1: &\max_{} \quad \frac{1}{M} \sum_{m=1}^{M}\sum_{j=1}^{J} l_{a_j,b_m} V_{b_m},   \\
    \mathrm{s.t.} \quad
    & \sum_{j=1}^{J}l_{a_j,b_m} = 1,  \label{sub-1}\\
    & \sum_{m=1}^{M}l_{a_j,b_m} \le 1,  \label{sub-5}\\
    & l_{a_j,b_m} T^{s}_{a_i, a_j} \geq T_{b_m}, \label{sub-3} \\
    & s \in \bm{S}_{a_j}, \label{sub-4} \\
    & {{  r_{a_i,a_j} \geq r_{b_m} }}. \label{sub-6}
\end{alignat}
Constraint (\ref{sub-1}) states that each subtask can be allocated to only one device, and constraint (\ref{sub-5}) stipulates that each device can be assigned at most one subtask. Constraint (\ref{sub-3}) states the task initiator's trust in a collaborator must meet the minimum trust demand of subtask $b_m$.
Constraint (\ref{sub-4}) stipulates that the prerequisite for assigning subtask $b_m$ with task type $s$ to device $a_j$ is that device $a_j$ must be capable of supporting tasks of task type $s$. Constraint (\ref{sub-6}) states that the transmission rate from device $a_i$ to device $a_j$ should meet the minimum transmission rate demand of subtask $b_m$.

\section{Task-specific Trusted Physical Resource Hypergraph}
\label{hypergraph-task-specific}

\begin{figure}[t!]
\centering
\includegraphics[scale=1]{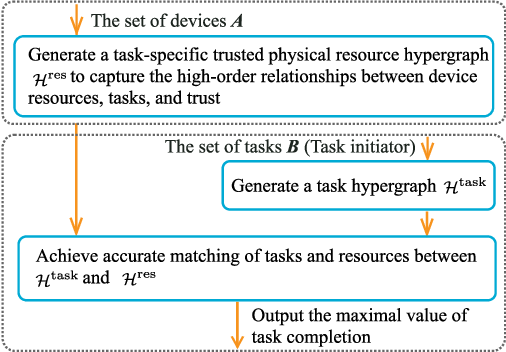}
\caption{The flowchart for implementing the TTR-matching model.}
\label{flowchat}
\end{figure}

To implement the proposed networked physical computing system, this study introduces an innovative TTR-matching model, which seamlessly integrates resources, tasks, and task-specific trust to enable value-maximizing task-resource matching. It consists of two key components, as shown in Fig~\ref{flowchat}. The first component constructs a task-specific trusted physical resource hypergraph $\mathcal{H}^{\text{res}}$, designed to achieve three objectives: evaluating task-specific trust, modeling collaboration dependencies among devices under specific tasks, and integrating the physical attributes of device resources. The second component consists of generating a task hypergraph, $\mathcal{H}^{\text{task}}$, that integrates the task initiator and the physical attributes of tasks, and performing precise task-resource matching between $\mathcal{H}^{\text{task}}$ and $\mathcal{H}^{\text{res}}$.

This section starts by presenting the fundamental concepts of hypergraphs, followed by the development of the task-specific trusted physical resource hypergraph.

\subsection{Hypergraph}

\begin{figure}[!t]
\centering
\includegraphics[scale=0.95]{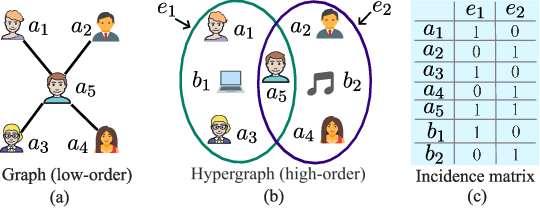}
\caption{Graph vs. hypergraph in representing collaborative relationships.}
\label{hypergraph}
\end{figure}

Hypergraphs have been widely applied due to their capability to accurately capture diverse relationships among entities in complex systems, offering greater expressive power than traditional graph structures. As shown in Fig.~\ref{hypergraph}, there is a social system consisting of five collaborators $\{a_1,a_2,a_3,a_4,a_5 \}$, who are capable of collaborating on various tasks, such as $b_1$ (coding) and $b_2$ (performing)\footnote{For the sake of symbol consistency, we use $a$ to represent collaborators and $b$ to represent tasks here.}. To enable effective collaboration, it is essential to accurately model the cooperative relationships among these collaborators. In Fig.~\ref{hypergraph}~(a), a graph can only represent cooperation through pairwise connections, which are low-order relationships. In contrast, Fig.~\ref{hypergraph}~(b) shows that each edge,  also known as a hyperedge, in a hypergraph can associate multiple collaborators with a specific task. For example, since $a_1$, $a_3$, and $a_5$ can collaborate to perform task $b_1$, hyperedge $e_1$ is used to link these three collaborators with task $b_1$, representing a high-order collaboration relationship. Therefore, the hypergraph that represents the collaborative relationships in this social system is denoted by $\mathcal{H} = (\mathcal{V}, \mathcal{E})$, where $\mathcal{V} $ is the set of all entities involved in the hypergraph and $\mathcal{E}$ is the set of all hyperedges.
$\mathcal{H}$ is also referred to as a $K^{\text{ord}}$-uniform hypergraph, where $K^{\text{ord}} = 4$ in this example, as each hyperedge includes four entities.
Typically, the incidence matrix $\mathcal{H} \in \mathbb{R}^{|\mathcal{V}| \times |\mathcal{E}|}$, as shown in Fig.~\ref{hypergraph}~(c), is employed to depict a hypergraph, which is given by \cite{10546264}
   \begin{align}
    \mathcal{H}(i,j) = \begin{cases}
    1, & \text{if} \ a_i \, \text{or} \, b_i \in e_j; \\
    0, & \text{otherwise}. \\
   \end{cases}
   \end{align}
Furthermore, a weighted hypergraph is commonly expressed as $\mathcal{H}=(\mathcal{V}, \mathcal{E}, \mathcal{W}^{\mathcal{V}}, \mathcal{W}^{\mathcal{E}})$, where $\mathcal{W}^{\mathcal{V}}$ and $\mathcal{W}^{\mathcal{E}}$ denote the weight parameters of entities and hyperedges, respectively. These weight parameters can represent various properties.

\subsection{Task-Specific Trusted Physical Resource Hypergraph}

According to equation (\ref{task-specific}), the task-specific trusted physical resource hypergraph $\mathcal{H}^{\text{res}}$ is implemented through the following three steps.

\begin{figure*}[t!]
\centering
\includegraphics[scale=0.7]{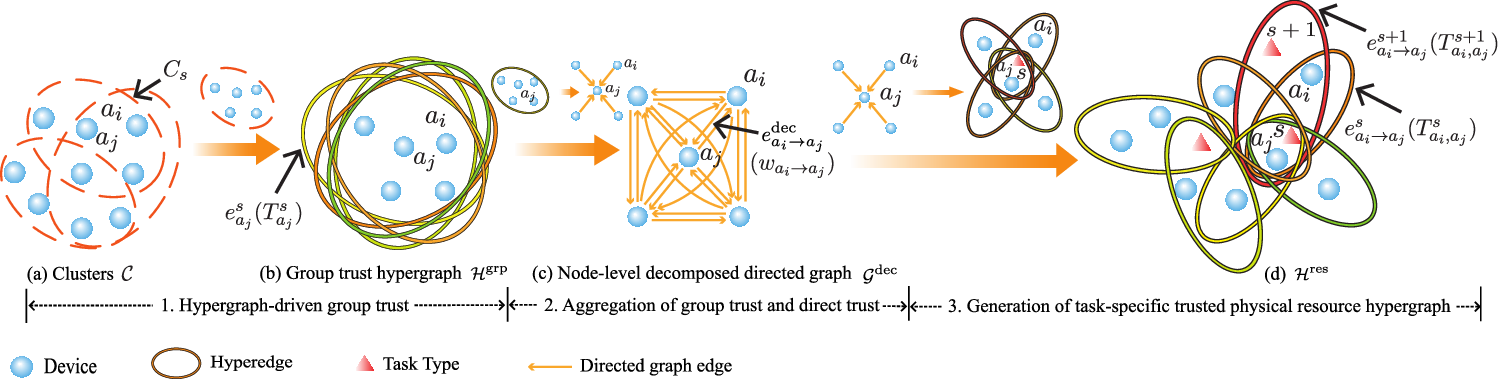}
\caption{The process of forming task-specific trusted physical resource hypergraph.}
\label{multi-tier hypergraph}
\end{figure*}

\textit{1) Hypergraph-driven group trust}: For any pair of devices, such as $a_i$ and $a_j$, the trust value of device $a_i$ towards device $a_j$ includes direct trust and indirect trust. Direct trust refers to the trust value assessed by device $a_i$ based on interactions and other information between them. Indirect trust refers to the trust value assessed by other devices in the system towards device $a_j$, also denoted as group trust in this research. 
To calculate group trust, all devices are first grouped into different clusters based on task types, denoted as $\mathcal{\bm{C}} = \{\bm{C}_{1}, \dots, \bm{C}_{S}\}$, as illustrated in Fig.~\ref{multi-tier hypergraph}~(a)\footnote{
Each hypergraph or graph in Fig.~\ref{multi-tier hypergraph} is only used as a simple example to show the process of generating the task-specific trusted physical resource hypergraph.}. Each cluster is specifically represented as 
 \begin{align}
    \bm{C}_{s} = \{l^{s}_{a_j} a_j, j = 1,\dots, J \}, s \in \bm{S},
\end{align}
where $l^{s}_{a_j} =  1$ if device $a_j$ supports task type $s$, and $l^{s}_{a_j} = 0$ otherwise. Within $\bm{C}_{s}$, devices have potential cooperative relationships to execute tasks with task type $s$. As device $a_j$ may support multiple task types, the devices that potentially cooperate with $a_j$ are represented as $\bm{G}_{a_j} = \bigcup_{s=1}^S l^s_{a_j}\bm{C}_{s}$. 


Then, $\mathcal{\bm{C}}$ is transformed into a group trust hypergraph. For $\bm{C}_s$, each device can form a hyperedge connecting to other devices in $\bm{C}_s$. The weight of this hyperedge is the group trust value assigned to this device. As shown in Fig.~\ref{multi-tier hypergraph}~(b), device $a_j$ generates a hyperedge $e^s_{a_j}$ that connects to all other devices in $\bm{C}_{s}$. These devices collectively determine the group trust value for $a_j$, which is calculated as 
\begin{align}\footnotesize
    T^{s}_{a_j} = \sum\limits_{|e^s_{a_j}|,i \neq j}^{}R(a_i,a_j), \ \ a_i,a_j \in e^s_{a_j},
\end{align}
where $|e^s_{a_j}|$ is the number of devices in $e^s_{a_j}$, and $R(a_i, a_j)$ represents the trust evaluation of device $a_i$ toward device $a_j$. $R(a_i, a_j)$ is calculated by taking into account their degree of cooperativeness, the proximity of their relationship, and their historical interactions, which is expressed as
\begin{align}
  \label{r}
  \hspace{-3mm} R(a_i,a_j) = \delta_1 \frac{|\bm{S}_{a_i} \! \cap \! \bm{S}_{a_j}|}{|\bm{S}_{a_i}\! \cup \!\bm{S}_{a_j}|}\! +\! \delta_2 \frac{|\bm{G}_{a_i} \!\cap\! \bm{G}_{a_j}|}{|\bm{G}_{a_i} \!\cup \!\bm{G}_{a_j}|}\! +\! \delta_3 \frac{\sum_{k=1}^{K_{a_i,a_j}} b_k^{\text{ret}} }{K_{a_i,a_j}},
\end{align}
where $\delta_1$, $\delta_2$, and $\delta_3$ are the weight parameters, $\delta_1 + \delta_2 + \delta_3 = 1$, $0 \le \delta_1, \delta_2, \delta_3 \le 1$. The first term on the right-hand side represents the degree of cooperativeness between two devices. The more they mutually support the same task types, the higher their degree of cooperativeness. The second term reflects the proximity of their relationship. The greater the number of common potential collaborators they have, the closer their relationship becomes. The third term represents the task completion rate of device $a_j$, which is calculated as the ratio of successful subtasks to the total number of subtasks. $K_{a_i,a_j}$ denotes the total number of historical subtasks assigned by device $a_i$ to device $a_j$. $b_k^{\text{ret}}$ represents the completion result of a subtask, which is calculated as 
\begin{align}
    b_k^{\text{ret}} = b^{\text{tra}}b^{\text{exe}},
\end{align}
where $b^{\text{tra}}$ and $b^{\text{exe}}$ are the transmission result and the execution result, respectively. $b^{\text{tra}}$ represents the success or failure of transmitting this subtask from $a_i$ to $a_j$, expressed as 
  \begin{align}
    b^{\text{tra}} = \begin{cases}
    1, &  \eta^{\text{pkt}} \le \eta^{\text{thr}},  \\
    0, & \text{otherwise}, \\
   \end{cases}
   \end{align}
where $\eta^{\text{pkt}}$ is the packet loss rate, and $\eta^{\text{thr}}$ is the threshold of the  packet loss rate. When $\eta^{\text{pkt}}$ is less than or equal to the threshold, the subtask transmission is considered successful~\cite{8727478}. $b^{\text{exe}}$ indicates the outcome of executing this subtask, which is given by
\begin{align}
    b^{\text{exe}} = \begin{cases}
    1, &  \text{success},  \\
    0, & \text{failure}. \\
   \end{cases}
\end{align}

\begin{algorithm}[t!]
	\caption{{Generation of the group trust hypergraph}}
	\label{generate implicit trust hypergraph}
	\begin{algorithmic}[1]
		\renewcommand{\algorithmicrequire}{\textbf{Input:}}
		\REQUIRE $\bm{A}$
            \renewcommand{\algorithmicrequire}{\textbf{Output:}}
		\REQUIRE Hypergraph $\mathcal{H}^{\text{grp}}$
             \STATE $\mathcal{H}^{\text{grp}} \xleftarrow[]{} \emptyset$, $\mathcal{\bm{C}} \xleftarrow[]{} \emptyset$, $\bm{C}_{s} \xleftarrow[]{} \emptyset$
             \FOR{$s$ in $\bm{S}$}
                 \FOR{$a_i$ in $\bm{A}$}
                     \IF{$l_{a_i}^{s} = 1$}
                     \STATE $\bm{C}_{s} \xleftarrow[]{add} a_i$
                     \ENDIF
                 \ENDFOR
                 \STATE $\mathcal{\bm{C}} \xleftarrow[]{add} \bm{C}_{s}$
             \ENDFOR
             \FOR{$\bm{C}_{s}$ in $\mathcal{\bm{C}}$}
                \FOR{each pair of $a_i$ and $a_j$ in $\bm{C}_{s}$}
                   \STATE calculating $R(a_i, a_j)$
                   \STATE calculating $T^{s}_{a_j}$
                   \STATE forming $e^s_{a_j}$ with the weight $T^{s}_{a_j}$
                   \STATE $\mathcal{H}^{\text{grp}} \xleftarrow[]{add} e^s_{a_j}$
                \ENDFOR
             \ENDFOR
             \RETURN $\mathcal{H}^{\text{grp}}$
	\end{algorithmic}
    \end{algorithm}

In the example shown in Fig.~\ref{multi-tier hypergraph}~(b),  $\bm{C}_s$ generates five hyperedges because it contains five devices. Ultimately, the hyperedges generated by devices in all clusters collectively form the group trust hypergraph $\mathcal{H}^{\text{grp}} = (\mathcal{V}^{\text{grp}}, \mathcal{E}^{\text{grp}}, \mathcal{W}^{\mathcal{V}^{\text{grp}}}, \mathcal{W}^{\mathcal{E}^{\text{grp}}})$ where $\mathcal{V}^{\text{grp}}$ is the set of all devices, $\mathcal{E}^{\text{grp}}$ is the set of all hyperedges, $\mathcal{W}^{\mathcal{V}^{\text{grp}}}$ represents the set of weights of devices with respect to each hyperedge, and $\mathcal{W}^{\mathcal{E}^{\text{grp}}}$ denotes the set of weights of hyperedges. $\mathcal{H}^{\text{grp}}$ leverages the advantages of hypergraphs to perfectly represent the indirect trust evaluation relationships between each device and its recommenders. The implementation of $\mathcal{H}^{\text{grp}}$ is presented in Algorithm \ref{generate implicit trust hypergraph}.

\textit{2) Aggregation of group trust and direct trust}: At this step, direct trust and group trust between devices need to be aggregated. To precisely represent the directions and weights of trust relationships, the node-level decomposition technique is used to convert $\mathcal{H}^{\text{grp}}$ into a weighted directed graph $\mathcal{G}^{\text{dec}}$.

Take a hyperedge $e^s_{a_j}$ in $\mathcal{H}^{\text{grp}}$ as an example. Since device $a_j$ is the central node of $e^s_{a_j}$, each device $a_i$, $a_i \in e^s_{a_j}$, $a_i \neq a_j$, can establish a directed edge pointing to device $a_j$, represented as $e^{\text{dec}}_{a_i \to a_j}$. By decomposing all hyperedges, the obtained $\mathcal{G}^{\text{dec}}$ is expressed as $(\mathcal{V}^{\text{dec}}, \mathcal{E}^{\text{dec}}, \mathcal{W}^{\mathcal{E}^{\text{dec}}})$, where $\mathcal{V}^{\text{dec}}$ is equal to $\mathcal{V}^{\text{grp}}$, $\mathcal{E}^{\text{dec}} = \{e^{\text{dec}}_{a_i \to a_j} = (a_i \to a_j): a_i, a_j \in e^s_{a_j}, i \neq j\} $ is the set of directed edges, and $\mathcal{W}^{\mathcal{E}^{\text{dec}}} = \{ w_{a_i \to a_j}\}$ is the set of weights of directed edges. $w_{a_i \to a_j}$ is the weight of the directed edge $e^{\text{dec}}_{a_i \to a_j}$, calculated as the weighted sum of overall trust and direct trust
\begin{align}
    \label{19}
    w_{a_i \to a_j} & = \beta_1 T_{a_j} + \beta_2 T_{a_i,a_j} \\ \nonumber
     & = \beta_1 \frac{\sum_{s=1}^{S} T^{s}_{a_j}}{\sum_{s=1}^{S}l^{s}_{a_j}} + \beta_2 R(a_i, a_j).
\end{align}
For the direct trust of device $a_i$ to device $a_j$, we denote its value as $R(a_i, a_j)$.
Since device $a_j$ may belong to multiple clusters, the overall trust $T_{a_j}$ is computed as the average of the group trust values from all clusters containing device $a_j$.

\textit{3) Generation of task-specific trusted physical resource hypergraph}: While the weighted directed graph $\mathcal{G}^{\text{dec}}$ effectively illustrates the directions and weights of trust relationships between devices, it falls short in capturing task-specific collaborations between devices. This is because a task-specific collaboration needs to involve four elements: two devices, a task type, and a trust relationship. Hence, a hypergraph is employed to integrate devices, tasks, and trust relationships to represent all possible task-specific collaborations. This hypergraph is called the task-specific trusted physical resource hypergraph, which is formalized as follows.

\begin{definition}[Task-specific trusted physical resource hypergraph] 
\textit{The task-specific trusted physical resource hypergraph is a hypergraph that accurately represents the interdependencies between tasks, physical resources, and trust in the networked physical computing system. It is defined as $\mathcal{H}^{\text{res}} = (\mathcal{V}^{\text{res}}, \mathcal{E}^{\text{res}}, \mathcal{W}^{\mathcal{V}^{\text{res}}}, \mathcal{W}^{\mathcal{E}^{\text{res}}})$, where $\mathcal{V}^{\text{res}}$ represents the set of all involved objects, including devices and task types, $\mathcal{E}^{\text{res}}$ is the set of all potential collaborations, $\mathcal{W}^{\mathcal{V}^{\text{res}}}$ is the physical attributes of all devices, and $\mathcal{W}^{\mathcal{E}^{\text{res}}}$ is the set of weights of all hyperedges. Each hyperedge $e^{s}_{a_i \to a_j} = (a_i,s,a_j) \in \mathcal{E}^{\text{res}}$ represents that device $a_i$ has a collaboration relationship, specific to task type $s$, directed towards device $a_j$, its weight is the trust value from  device $a_i$ to device $a_j$ specific to task type $s$. }
\end{definition}

\begin{figure}[t!]
\centering
\includegraphics[scale=0.75]{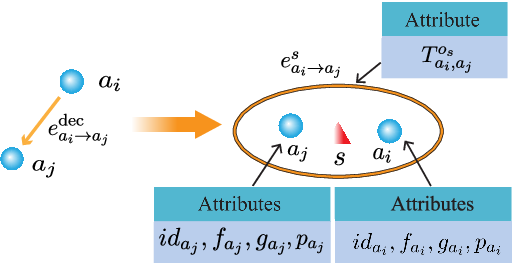}
\caption{Transforming the directed edge $e^{\text{dec}}_{a_i \to a_j}$ into the hyperedge $e^{s}_{a_i \to a_j}$.}
\label{hyperedge}
\end{figure}

To obtain $\mathcal{H}^{\text{res}}$, each directed edge $e^{\text{dec}}_{a_i \to a_j}$ in $\mathcal{G}^{\text{dec}}$ is transformed into a hyperedge $e^{s}_{a_i \to a_j}$, as shown in Fig.~\ref{hyperedge}. The attributes of each device $a_i$ in $e^{s}_{a_i \to a_j}$ include $id_{a_i}$, $f_{a_i}$, and $g_{a_i}$, and the weight $w^{s}_{a_i \to a_j}$ of $e^{s}_{a_i \to a_j}$ is the trust value of $a_i$ to $a_j$ specific to task type $s$, which is calculated by
\begin{align}
  \label{20}
    T^{s}_{a_i,a_j} = w_{a_i \to a_j} + \beta_3 T_{a_i,s,a_j},
\end{align}
where
\begin{align}
    T_{a_i,s,a_j} = \frac{\sum_{k=1}^{K^s_{a_i,a_j}}b_k^{\text{ret}}}{K^s_{a_i,a_j}},
\end{align}
where ${K^s_{a_i,a_j}}$ is the total number of historical subtasks of task type $s$ assigned by device $a_i$ to device $a_j$.

Finally, all the obtained hyperedges form the task-specific trusted physical resource hypergraph $\mathcal{H}^{\text{res}} = (\mathcal{V}^{\text{res}}, \mathcal{E}^{\text{res}}, \mathcal{W}^{\mathcal{V}^{\text{res}}}, \mathcal{W}^{\mathcal{E}^{\text{res}}})$. It is important to note that when two devices support multiple task types, there are multiple task-specific trust relationships between them. As shown in Fig.~\ref{multi-tier hypergraph}~(d), since both device $a_i$ and device $a_j$ support task types $s$ and $s+1$, device $a_i$ can build two hyperedges pointing to device $a_j$.  
One hyperedge, $e^s_{a_i \to a_j}$, with weight $T^s_{a_i,a_j}$, is specific to task type $s$, while another, $e^{s+1}_{a_i \to a_j}$, with weight $T^{s+1}_{a_i,a_j}$, is specific to task type $s+1$.
The generation of $\mathcal{H}^{\text{res}}$ is outlined in Algorithm~\ref{generate multi tier hypergraph} and illustrated in Fig.~\ref{multi-tier hypergraph}.  

\begin{algorithm}[t!]
	\caption{{Generation of the task-specific trusted physical resource hypergraph}}
	\label{generate multi tier hypergraph}
	\begin{algorithmic}[1]
		\renewcommand{\algorithmicrequire}{\textbf{Input:}}
		\REQUIRE Hypergraph $\mathcal{H}^{\text{grp}}$
            \renewcommand{\algorithmicrequire}{\textbf{Output:}}
		\REQUIRE Hypergraph $\mathcal{H}^{\text{res}}$
             \STATE $\mathcal{G}^{\text{dec}} \xleftarrow[]{} \emptyset$, $\mathcal{H}^{\text{res}} \xleftarrow[]{} \emptyset$
             
             \FOR{$e^s_{a_j}$ in $\mathcal{H}^{\text{grp}}$}
                \FOR{$a_i$ in  $e^s_{a_j}, a_i \neq a_j$}
                    \STATE creating the directed graph edge $e^{\text{dec}}_{a_i \to a_j}$ 
                    \STATE calculating $w_{a_i \to a_j}$ by (\ref{19})
                    \STATE $\mathcal{G}^{\text{dec}} \xleftarrow[]{add} e^{\text{dec}}_{a_i \to a_j}$
                \ENDFOR
             \ENDFOR
             \FOR{$e^{\text{dec}}_{a_i \to a_j}$ in 
                  $\mathcal{G}^{\text{dec}}$}
                   \STATE transforming $e^{\text{dec}}_{a_i \to a_j}$ into $e^s_{a_i \to a_j}$
                   \STATE calculating $w^{s}_{a_i \to a_j}$ by (\ref{20})
                   \STATE $\mathcal{H}^{\text{res}} \xleftarrow[]{add} e^{s}_{a_i \to a_j}$
             \ENDFOR
             \RETURN $\mathcal{H}^{\text{res}} = (\mathcal{V}^{\text{res}}, \mathcal{E}^{\text{res}}, \mathcal{W}^{\mathcal{V}^{\text{res}}}, \mathcal{W}^{\mathcal{E}^{\text{res}}})$
	\end{algorithmic}
 \end{algorithm}

$\mathcal{H}^{\text{res}}$ plays a critical role in constructing the networked physical computing system.  By enabling task-specific trust evaluation, it enhances the precision of collaborator selection. Furthermore, it provides a clearer representation of collaboration dependencies among devices, capturing how their interactions vary under different tasks, which facilitates better coordination and resource utilization. Lastly, by integrating the diverse physical attributes of device resources, $\mathcal{H}^{\text{res}}$ effectively accounts for heterogeneity in device capabilities, thereby enhancing the system’s ability to match tasks with suitable resources.

\section{Task-Resource Matching}
\label{section_hypergraph_matching}
In the previous section, the task-specific trusted physical resource hypergraph is proposed to integrate device resources, tasks, and task-specific trust. This section focuses on implementing the precise matching of tasks and resources to maximize the value of task completion. First, the generation of the task hypergraph $\mathcal{H}^{\text{task}}$ will be introduced. Then, a hypergraph matching algorithm will be employed to achieve the matching between $\mathcal{H}^{\text{task}}$ and $\mathcal{H}^{\text{res}}$.

\subsection{Task Hypergraph} 
To distinguish from symbols used in $\mathcal{H}^{\text{res}}$, we use symbols $a'_i$ and $s'$ to represent the task initiator and task type in the task hypergraph, respectively. For a task $\bm{B}$ generated by the task initiator $a_i'$, its task hypergraph is denoted as $\mathcal{H}^{\text{task}} = (\mathcal{V}^{\text{task}}, \mathcal{E}^{\text{task}}, \mathcal{W}^{\mathcal{V}^{\text{task}}}, \mathcal{W}^{\mathcal{E}^{\text{task}}})$ where $\mathcal{V}^{\text{task}}$ is the set of all involved objects, including all subtasks and $a_i'$, $\mathcal{E}^{\text{task}}$ is the set of all task hyperedges, $\mathcal{W}^{\mathcal{V}^{\text{task}}}$ is the collection of attributes of all objects, and $\mathcal{W}^{\mathcal{E}^{\text{task}}}$ is the set of weights of all hyperedges. The attributes of subtasks are defined in Section~\ref{Modeling of Trusted Collaborative IoT system}-A. Each task hyperedge $e^{s'}_{a_i' \to \phi} = (a_i', s', \phi) \in \mathcal{E}^{\text{task}}$ connects $a_i'$, a subtask $b_m \in \bm{B}$ with task type $s' \in \bm{S}$, and $\phi$. 
$\phi$ is the virtual node without practical meaning. The weight $w^{s'}_{a_i' \to \phi} \in \mathcal{W}^{\mathcal{E}^{\text{task}}}$ of $e^{s'}_{a_i' \to \phi}$ is set to $T_{b_m}$, which is the minimum trust demand of subtask $b_m$. The generation of the task hypergraph $\mathcal{H}^{\text{task}}$ is presented in Algorithm~\ref{generate task hypergraph}.

\begin{algorithm}[t!]
	\caption{{Generation of the task hypergraph}}
	\label{generate task hypergraph}
	\begin{algorithmic}[1]
		\renewcommand{\algorithmicrequire}{\textbf{Input:}}
		\REQUIRE $\bm{B}$, $a_i'$
            \renewcommand{\algorithmicrequire}{\textbf{Output:}}
		\REQUIRE $\mathcal{H}^{\text{task}}$
             \STATE $\mathcal{H}^{\text{task}} \xleftarrow[]{} \emptyset$
              \FOR{$b_m$ in $\bm{B}$}
                  \STATE building the hyperedge $e^{s'}_{a_i' \to \phi}$ between $a_i'$ and $b_m$ of task type $s'$
                  \STATE associating $T_{b_m}$ to $e^{s'}_{a_i' \to \phi}$ as its weight
                  \STATE $\mathcal{H}^{\text{task}} \xleftarrow[]{add} e^{s'}_{a_i' \to \phi}$
              \ENDFOR
             \RETURN $\mathcal{H}^{\text{task}} = (\mathcal{V}^{\text{task}}, \mathcal{E}^{\text{task}}, \mathcal{W}^{\mathcal{V}^{\text{task}}}, \mathcal{W}^{\mathcal{E}^{\text{task}}})$
	\end{algorithmic}
 \end{algorithm}

\subsection{Matching of $\mathcal{H}^{\text{res}}$ and $\mathcal{H}^{\text{task}}$}

The goal of hypergraph matching is to find correspondences between two hypergraphs by considering the affinities of their corresponding vertices and hyperedges such that the matching score is maximized. Take the matching between a task hyperedge $e^{s'}_{a_i' \to \phi} = (a_i', s', \phi)$ from $\mathcal{H}^{\text{task}}$ and a resource hyperedge $e^{s}_{a_i \to a_j} = (a_i, s, a_j)$ from $\mathcal{H}^{\text{res}}$ as an example. A tuple $(f_{a_i'a_i},f_{s' s},f_{\phi a_j})$ is used to represent hyperedge correspondence, with each $f$ indicating a one-to-one mapping between two vertices. If $a_i'$ matches $a_i$, i.e., they are the same device, $f_{a_i' a_i} = 1$, otherwise $f_{a_i' a_i} = 0$. Similarly, if $s'$ matches $s$, $f_{s' s} = 1$, otherwise $f_{s' s} = 0$. If $f_{a_i'a_i} = 1, f_{s' s} = 1$, and $w^{s}_{a_i \to a_j} \geq w^{s'}_{a_i \to \phi}$, i.e., the trust of $a_i$ in $a_j$ specific to $s$ is greater than the trust demand of $b_m$, then $a_j$ is considered a potential collaborative device to execute $b_m$ and $f_{\phi a_j}$ is set to 1. The matching score between $e^{s'}_{a_i' \to \phi}$ and $e^{s}_{a_i \to a_j}$ can be calculated by equation (\ref{value}), and the true match is accompanied by a large matching score.

As hypergraph matching aims to maximize the overall matching score between two hypergraphs, it is essential to identify the mapping between vertices in these two hypergraphs. We assume that the number of objects in hypergraph $\mathcal{H}^{\text{task}}$ is significantly smaller than the number of objects in hypergraph $\mathcal{H}^{\text{res}}$, i.e., $|\mathcal{V}^{\text{task}}| < |\mathcal{V}^{\text{res}}|$. The set of all possible assignment matrices that assign each vertex of $\mathcal{H}^{\text{task}}$ to exactly one vertex of $\mathcal{H}^{\text{res}}$ is expressed as a binary assignment matrix $\bm{f} \in \{0, 1\}^{\mathcal{V}^{\text{task}} \times \mathcal{V}^{\text{res}}}$. To achieve the optimal matching from $\mathcal{H}^{\text{task}}$ to $\mathcal{H}^{\text{res}}$, we aim to maximize the the matching score
\begin{alignat}{1}
    P2: &\max_{\bm{f}} \quad \sum^{\mathcal{V}^{\text{task}}}_{a'_i,s',\phi} \sum^{\mathcal{V}^{\text{res}}}_{a_i,s,a_j} F f_{a_i'a_i} f_{s' s} f_{\phi a_j}, 
\end{alignat}
where $F$ denotes the function used to calculate the matching score between each pair of task hyperedge $e^{s'}_{a'_i \to \phi}$ and resource hyperedge $e^{s}_{a_i \to a_j}$. In this research, equation~(\ref{value}) is adopted as the implementation of $F$. The product $f_{a_i'a_i} f_{s' s} f_{\phi a_j}$ equals 1 if and only if $e^{s'}_{a'_i \to \phi} = (a_i',s',\phi)$ matches $e^{s}_{a_i \to a_j} = (a_i,s,a_j)$. One can see that the optimization problem $P1$ is effectively transformed into the hypergraph matching problem $P2$ through the integration of physical device resources, task-specific trust, and tasks.

\subsection{Hypergraph Matching as a Clustering Game}

Existing methods for solving the hypergraph matching problem are primarily approximation-based and exhibit high computational complexity \cite{7858754}. In this research, the hypergraph matching problem is considered as a non-cooperative multiplayer clustering game where the correct matches between $\mathcal{H}^{\text{task}}$ and $\mathcal{H}^{\text{res}}$ can form a related group that receives the highest matching score values. 

\begin{figure}[t!]
\centering
\includegraphics[scale=0.75]{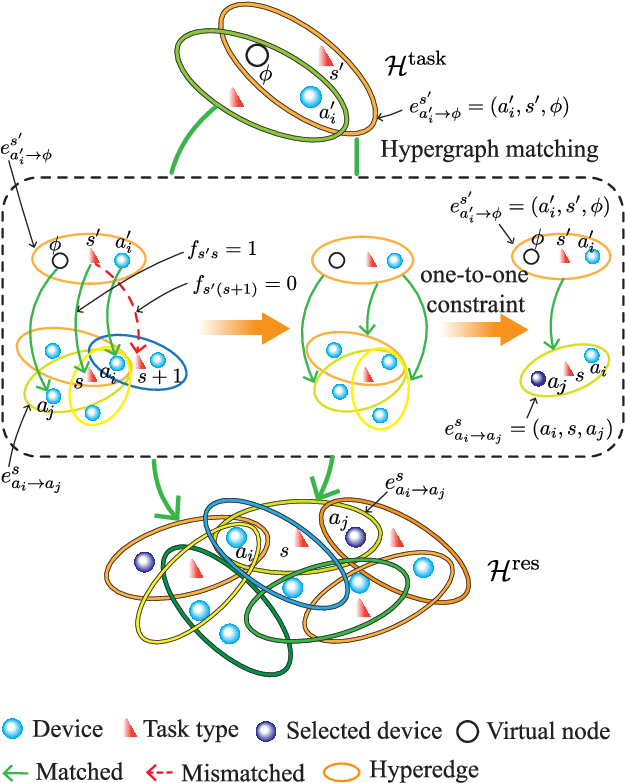}
\caption{TTR-matching model.}
\label{hypergraph matching}
\end{figure}

In terms of game theory \cite{8545827}, a game of strategy is formalized as $\Gamma = (\bm{P}, \bm{N}, \pi)$, where $\bm{P}=\{1,\dots, P\}$ is the set of $P$ players, $\bm{N}$ is the set of pure strategies that are available to players. $\pi: \bm{N}^{P} \rightarrow \mathbb{R}$ is the payoff function that assigns a payoff to each strategy profile defined as an ordered group of strategies played by the different players. If hypergraph matching considers all point-to-point connections between the two hypergraphs as strategies, the computational complexity will become unmanageable. To reduce complexity, in this research, a candidate match between a task hyperedge $e^{s'}_{a'_i \to \phi}$ and a resource hyperedge $e^{s}_{a_i \to a_j}$ is considered as a strategy. Hence, numerous coarse correspondences between the task hyperedge set $\mathcal{E}^{\text{task}}$ and the resource hyperedge set $\mathcal{E}^{\text{res}}$ are constructed and regarded as pure strategies $\bm{N} = \{1, \dots, N\}$. The matching score values are considered as the payoffs of the players. The clustering game is played in an evolutionary scenario where three players are drawn from a large population to repeatedly play the game. These players are not assumed to act rationally or possess comprehensive knowledge of the game details. The set of all states of the population is represented by the standard simplex $\Delta$ of $N$-dimensional Euclidean space
\begin{align}
    \setlength\abovedisplayskip{2pt}
    \hspace{-2.5mm} \Delta = \{\bm{q} \in \mathbb{R}^N : \sum_{n=1}^{N}q_n=1 \ \text{and} \ q_n \geq 0 \ \text{for all}\ n \in \bm{N}\},
    \setlength\belowdisplayskip{2pt}
\end{align}
where $q_n$ is the fraction of players selecting the $n$-th strategy. Over time, the population undergoes changes in states driven by the influence of the selection process. Following Darwinian Natural Selection theory, players who achieve substantial payoffs thrive over evolution, while those with lower payoffs gradually diminish. As payoffs are evaluated by the matching score values, it is evident that survived strategies (matches) played by survived players receive high matching score values, while extinct ones are associated with lower matching score values. Ultimately, the evolutionary process reaches an equilibrium $\bm{q} \in \Delta$ when each survived player obtains the same expected payoff, i.e., $\bm{q} \in \Delta$ is a Nash equilibrium if 
\begin{align}
    u\left( \bm{\theta}^n, \bm{q}^{[2]} \right) \leq u\left(\bm{q}^{[3]}\right),\ \text{for all} \ n \in \bm{N},  
\end{align}
where the left side signifies the expected payoff acquired by players choosing the $n$-th strategy, while the right side is the expected payoff over the entire population. $\bm{\theta}^{n}$ represents an $N$-dimensional state vector with $q_n=1$ and zero elsewhere. $\bm{q}^{[3]}$ and $\bm{q}^{[2]}$ stand for 3 and 2 identical states $\bm{q}$, respectively, and the function $u$ is defined as follows
\begin{align}
    u \left( \bm{h}^{(1)}, \bm{h}^{(2)}, \bm{h}^{(3)} \right) = \!\sum_{(n_1,n_2,n_3)\in \bm{N}^3}\! \pi \left(n_1,n_2,n_3 \right) \prod_{p=1}^{3} h^{(p)}_{n_p},
\end{align}
where $\pi(n_1,n_2,n_3)$ represents the assigned payoff to the strategy profile $(n_2,n_2,n_3)$ and $n_p$ is the strategy selected by the $p$-th player. 

According to the conclusion in \cite{6330964}, the survived strategies (matches) constitute a consistent group which can be obtained by finding out the evolutionary stably strategy (ESS) cluster of the game. Based on the Baum-Eagon inequality, the ESS cluster can be determined by \cite{6330964}
\begin{align}
    \label{ess}
     q_n(t^{\text{gam}} + 1) = q_n(t)\frac{ u\left( \bm{\theta}^n, \bm{q}(t^{\text{gam}})^{[2]} \right)}{u\left(\bm{q}(t^{\text{gam}})^{[3]}\right)}, n = 1,\dots, N.
\end{align}
At time $t^{\text{gam}}=0$, $\bm{q}$ is initialized with the initial state $\bm{q}(0)$. Then, it is updated iteratively using equation (\ref{ess}) until the final state $\bm{q}$ is obtained at convergence. The survived strategies that persist with weights exceeding a specified threshold in $\bm{q}$ form a cluster $\bm{C}^{\text{ESS}}$. Within this cluster, matches between $\mathcal{H}^{\text{task}}$ and $\mathcal{H}^{\text{res}}$ are regarded as correct matches. Each match is assigned a weight that denotes its degree of alignment with other matches. A higher weight serves as an indicator of a greater likelihood that the match is accurate. In instances of one-to-many matches within the cluster, the match with the highest weight is retained, adhering to the one-to-one constraint. Finally, the cluster $\bm{C}^{\text{match}}$ encapsulating the correct matches is obtained. As shown in Fig.~\ref{hypergraph matching}, there are three resource hyperedges that have potential matches with the task hyperedge $e^{s'}_{a'_i \to \phi}$. After the clustering game and one-to-one restriction, $e^s_{a_i \to a_j}$ and $e^{s'}_{a'_i \to \phi}$ are the best match. Then, $a_j$ in $e^s_{a_i \to a_j}$ is the selected collaborator to perform $b_m$ with task type $s$. The detailed implementation of the game-theoretic hypergraph matching algorithm is presented in Algorithm \ref{game matching}. Through the proposed networked physical computing system and the TTR-matching model, tasks and resources can be accurately matched to maximize the value of task completion.

 \begin{algorithm}[t!] 
	\caption{{Game-theoretic hypergraph matching algorithm}}
	\label{game matching}
	\begin{algorithmic}[1]
		\renewcommand{\algorithmicrequire}{\textbf{Input:}}
		\REQUIRE $\mathcal{H}^{\text{task}}, \mathcal{H}^{\text{res}}$,  the maximum number of iterations $\zeta$
            \renewcommand{\algorithmicrequire}{\textbf{Output:}}
		\REQUIRE $\bm{C}^{\text{match}}$ 
            \STATE $\bm{N} \leftarrow \emptyset$ \\
            \FOR{each $e^{s'}_{a'_i \to \phi}$ in $\mathcal{H}^{\text{task}}$}
               \STATE $\bm{N} \xleftarrow{add}$ coarse matches between $e^{s'}_{a'_i \to \phi}$ and hyperedges of $\mathcal{H^{\text{res}}}$
            \ENDFOR
            \STATE $t^{\text{gam}} \leftarrow 0$ \\
            \STATE $\bm{q} \leftarrow \bm{q}(0)$\\
            \WHILE{$t^{\text{gam}} < \zeta$}
                \STATE $t^{\text{gam}} \leftarrow t^{\text{gam}} + 1$ \\
                 \FOR{each $n$ in $\bm{N}$}
                     \STATE updating $q_n(t^{\text{gam}})$ using equation (\ref{ess})\\
                 \ENDFOR
                \STATE $\bm{q} \leftarrow \bm{q}(t^{\text{gam}})$\\
            \ENDWHILE
            \STATE $\bm{C}^{\text{ESS}} \leftarrow \bm{q}$
            \STATE $\bm{C}^{\text{match}} \xleftarrow[]{}{\bm{C}^{\text{ESS}}}$\\
		\RETURN  $\bm{C}^{\text{match}}$\\
	\end{algorithmic}
\end{algorithm}

\begin{figure*}[!t]
      \centering
      \subfigure[$a_1$ to $a_2$]{\includegraphics[scale=0.43]{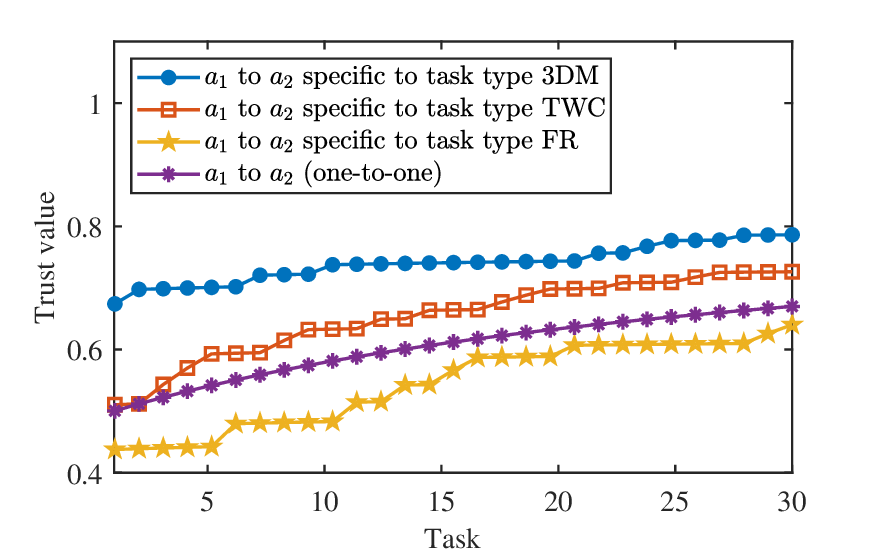}} \hspace{-0.29 in}
      \subfigure[ $a_1$ to $a_3$]{\includegraphics[scale=0.43]{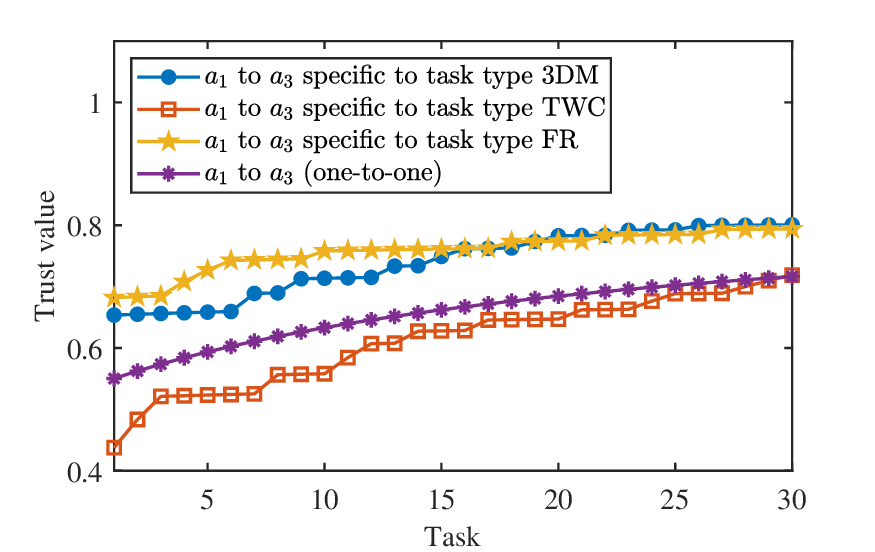}}
      \hspace{-0.29 in}
      \subfigure[ $a_1$ to $a_{16}$]{\includegraphics[scale=0.43]{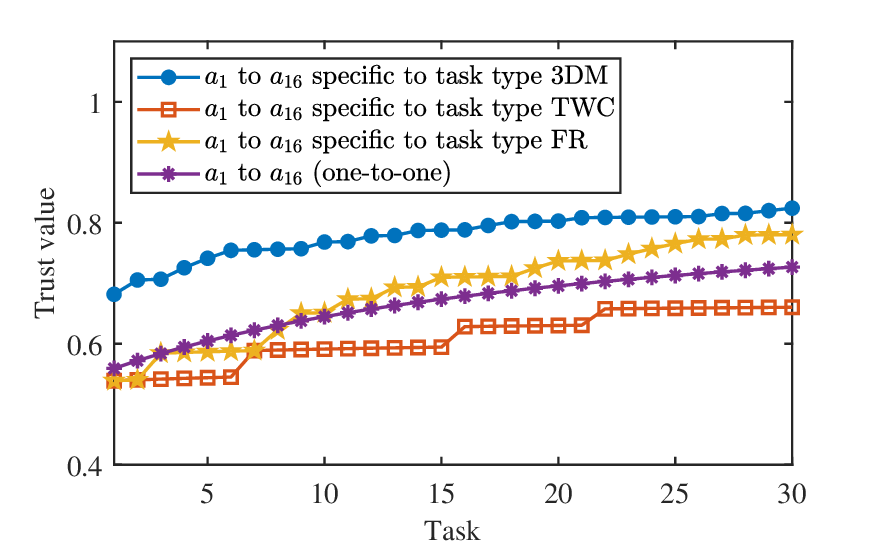}}
     \caption{Comparison of one-to-one trust and task-specific trust when $\beta_1 = 0.1$, $\beta_2 = 0.1$, and $\beta_3 = 0.8$.}
     \label{trust_comparison_0.1_0.1_0.8}
\end{figure*}

 \begin{figure*}[!t]
      \centering
      \subfigure[$a_1$ to $a_2$]{\includegraphics[scale=0.43]{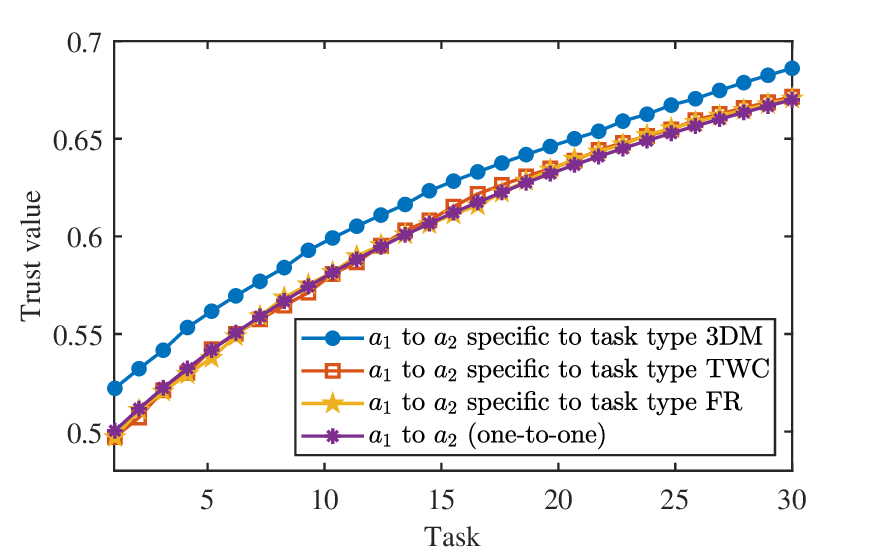}} \hspace{-0.29 in}
      \subfigure[ $a_1$ to $a_3$]{\includegraphics[scale=0.43]{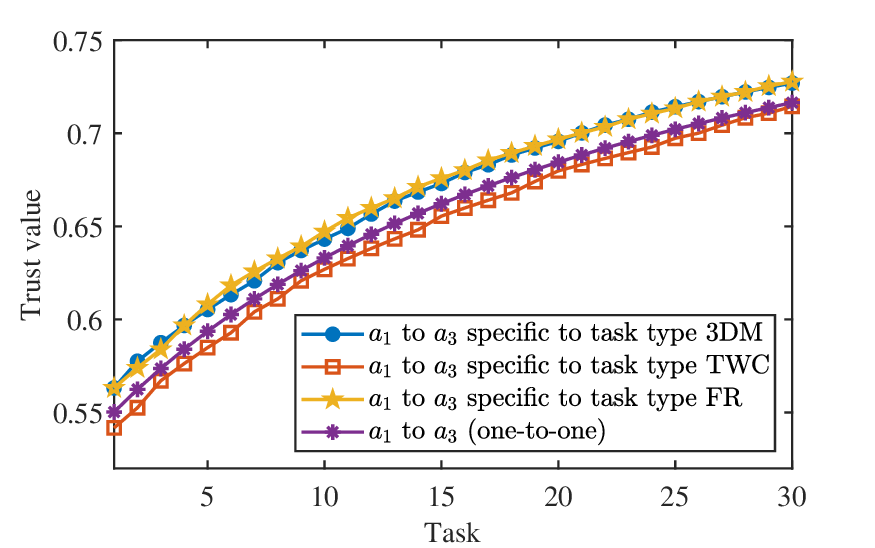}}
      \hspace{-0.29 in}
      \subfigure[ $a_1$ to $a_{16}$]{\includegraphics[scale=0.43]{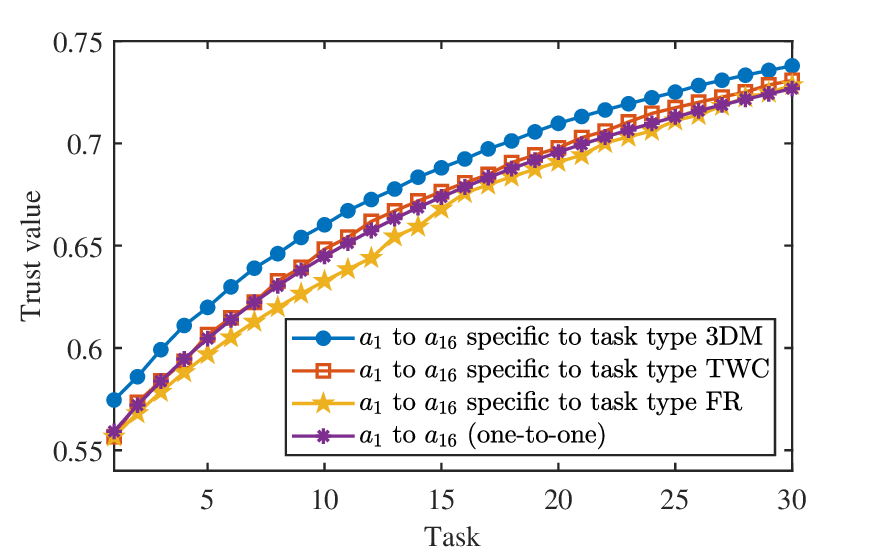}}
     \caption{Comparison of one-to-one trust and task-specific trust when $\beta_1 = 0.1$, $\beta_2 = 0.8$, and $\beta_3 = 0.1$.}
     \label{trust_comparison_0.1_0.8_0.1}
\end{figure*}

 \begin{figure*}[!t]
      \centering
      \subfigure[$a_1$ to $a_2$]{\includegraphics[scale=0.43]{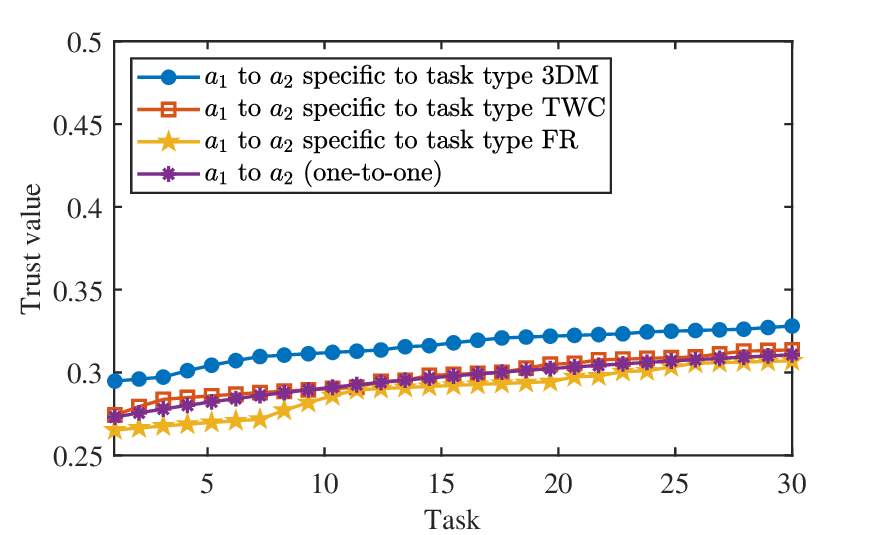}} \hspace{-0.29 in}
      \subfigure[ $a_1$ to $a_3$]{\includegraphics[scale=0.43]{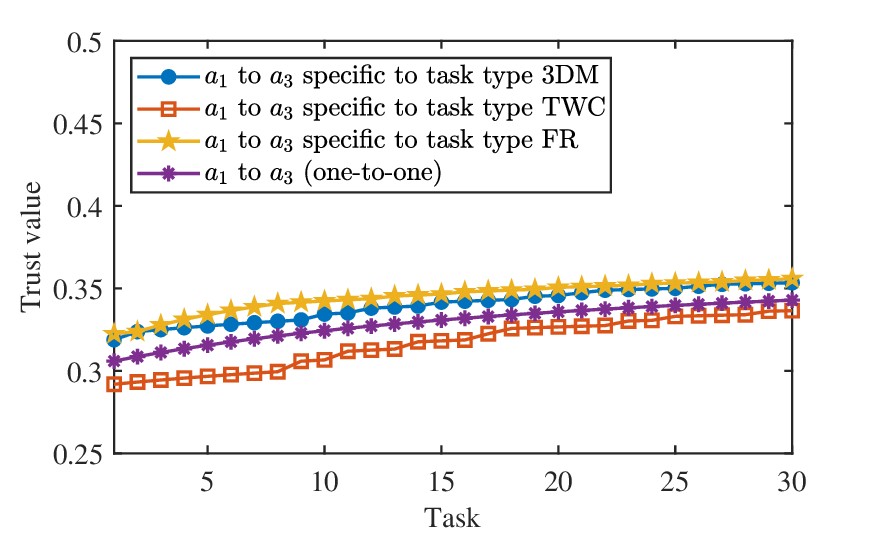}}
      \hspace{-0.29 in}
      \subfigure[ $a_1$ to $a_{16}$]{\includegraphics[scale=0.43]{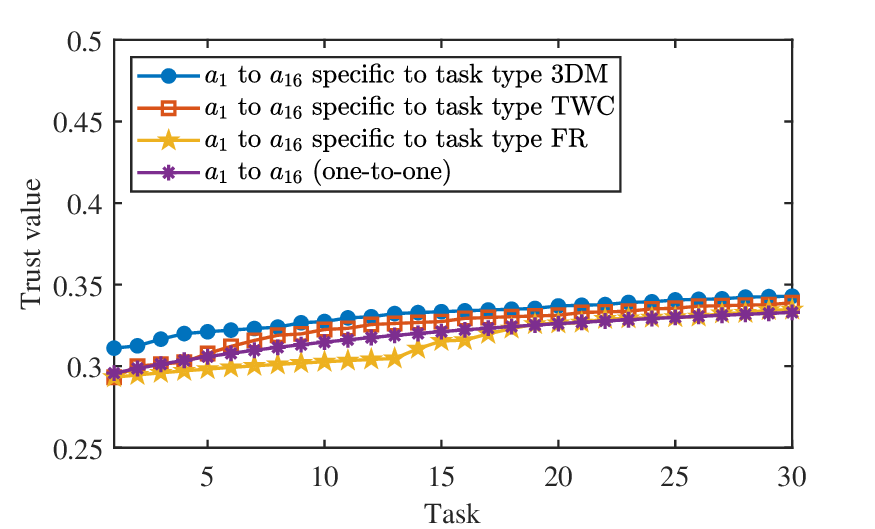}}
     \caption{Comparison of one-to-one trust and task-specific trust when $\beta_1 = 0.8$, $\beta_2 = 0.1$, and $\beta_3 = 0.1$.}
     \label{trust_comparison_0.8_0.1_0.1}
\end{figure*}

\subsection{Computational Complexity}
As stated in \cite{5871640}, when applying their proposed point-to-point hypergraph matching algorithms to match hypergraph $\mathcal{H}^{\text{task}} = (\mathcal{V}^{\text{task}}, \mathcal{E}^{\text{task}}, \mathcal{W}^{\mathcal{V}^{\text{task}}}, \mathcal{W}^{\mathcal{E}^{\text{task}}})$ and hypergraph $\mathcal{H}^{\text{res}} = (\mathcal{V}^{\text{res}}, \mathcal{E}^{\text{res}}, \mathcal{W}^{\mathcal{V}^{\text{res}}}, \mathcal{W}^{\mathcal{E}^{\text{res}}})$, the computational complexity of each iteration is $O(|\mathcal{V}^{\text{res}}|^{2K^{\text{ord}}})$, where $K^{\text{ord}} = 3$ as both $\mathcal{H}^{\text{task}}$ and $\mathcal{H}^{\text{res}}$ are 3-order uniform hypergraphs. However, to reduce computational complexity, our game-theoretic hypergraph matching approach treats each potential match between a task hyperedge and a resource hyperedge as an individual strategy. As a result, the computational complexity per iteration is $O(N)$, where $N$ is the number of strategies.

\section{Experimental Results and Discussion}
\label{simulation_discussion}
In this section, the proposed TTR-matching model is evaluated through numerical experiments conducted on the Intelligent Collaborative System (ICS) platform. The ICS platform is composed of 1 iPad ($a_1$), 8 Google pixel 8 phones ($a_2, a_3, a_4,a_5,a_6,a_7,a_8,a_9$), 3 DELL 5200 edge servers ($a_{10}, a_{11}, a_{12}$), 3 DELL precision 5820 desktops ($a_{13}, a_{14}, a_{15}$), 2 Lambda workstations ($a_{16}, a_{17}$), 4 Rosbot Plus robots ($a_{18}, a_{19}, a_{20}, a_{21}$), and 5 Robofleet robots ($a_{22}, a_{23}, a_{24}, a_{25}, a_{26}$). The parameters of these devices and the experimental setup are provided in Appendix~A. Following the descriptions of tasks provided in \cite{7264984}\cite{7091048}\cite{10.5555/1863103.1863107}\cite{9301339}\cite{6195845}, we consider that the ICS platform supports the following four task types: 1) face recognition (FR), with a processing density of 2,339 cycles/bit~\cite{7264984}, 2) video transcoding (VT), with a processing density of 1000 cycles/bit~\cite{7091048}, 3) text word count (TWC), with a processing density of 16,800 cycles/bit~\cite{10.5555/1863103.1863107}, and 4) 3D mapping (3DM), with a processing density of 1500 cycles/bit~\cite{9301339}.

Before running the proposed TTR-matching model, we first initialize the trust values of the devices in the system by executing 500 tasks. In each task, a device is randomly selected as the task initiator to generate a task containing multiple subtasks. Each subtask is assigned parameters, including size, task completion tolerance time, minimum trust and transmission rate requirements.

\subsection{Comparison of One-to-One Trust and Task-Specific Trust}
To assess the advantages of the proposed task-specific trust model, we compare it with the one-to-one trust model when varying the values of $\beta_1$, $\beta_2$, and $\beta_3$. Device $a_1$ serves as the task initiator to generate tasks, with the details of each task provided in Table~\ref{task}. Devices $a_2$, $a_3$, and $a_{16}$ are designated as collaborators, and $a_1$ randomly assigns the three subtasks to them in each task. Subsequently, the changes in the trust values of these three collaborators are observed after 30 tasks.

In Fig.~\ref{trust_comparison_0.1_0.1_0.8}, $\beta_1$, $\beta_2$, and $\beta_3$ are set to 0.1, 0.1, and 0.8, respectively, for calculating task-specific trust. For one-to-one trust, equation (\ref{task-specific}) is applied by setting $\beta_1=0.1$, $\beta_2 = 0.9$ and $\beta_3 = 0$, i.e., without considering $T_{a_i,s,a_j}$. We can see that for all collaborators, the task-specific trust values observed on them are significantly different from the one-to-one trust values. For example, as illustrated in Fig.~\ref{trust_comparison_0.1_0.1_0.8}~(a), after 30 tasks, the one-to-one trust value from device $a_1$ to device $a_2$ is 0.67. The task-specific trust values differ across task types: 0.79 for 3DM and 0.64 for FR. Suppose device $a_1$ has a new task composed of two subtasks: $b_1$ with a minimum trust threshold of 0.7, and $b_3$ with a minimum trust threshold of 0.67. If the one-to-one trust value is used, device $a_2$ would not be selected to perform subtask $b_1$ because its trust value (0.67) falls below the minimum trust threshold (0.7) required by $b_1$. However, when considering task-specific trust for the 3DM task type, device $a_2$’s trust value (0.79) exceeds the threshold, making it a suitable collaborator for $b_1$. Similarly, for subtask $b_3$, device $a_2$ would be selected as a collaborator based on its one-to-one trust value, which satisfies the minimum threshold of 0.67 required by subtask $b_3$. However, according to the task-specific trust for the FR task type, device $a_2$ would not qualify as a collaborator for subtask $b_3$ because its trust value (0.64) falls below the minimum threshold (0.67) of subtask $b_3$. This example demonstrates that task-specific trust offers a more precise and task-aware criterion for collaborator selection than general one-to-one trust. In Fig.~\ref{trust_comparison_0.1_0.8_0.1}, the parameters $\beta_1$, $\beta_2$, and $\beta_3$ are assigned values of 0.1, 0.8, and 0.1, respectively, for computing task-specific trust. For one-to-one trust, the values are set as $\beta_1=0.1$, $\beta_2=0.9$, and $\beta_3=0$. In Fig.~\ref{trust_comparison_0.8_0.1_0.1}, $\beta_1$, $\beta_2$, and $\beta_3$ are configured as 0.8, 0.1, and 0.1, respectively, for task-specific trust, while for one-to-one trust, the parameters are set to $\beta_1=0.8$, $\beta_2=0.2$, and $\beta_3=0$. 
The results observed in Fig.~\ref{trust_comparison_0.1_0.8_0.1} and Fig.~\ref{trust_comparison_0.8_0.1_0.1} are the same as those in Fig.~\ref{trust_comparison_0.1_0.1_0.8}, which demonstrates that the proposed task-specific trust model can effectively capture both the diversity and dynamism of trust relationships among devices.

\begin{figure*}[!t]
      \centering
      \subfigure[$\xi_1 = 0$, $\xi_2 = 1$]{\includegraphics[scale=0.43]{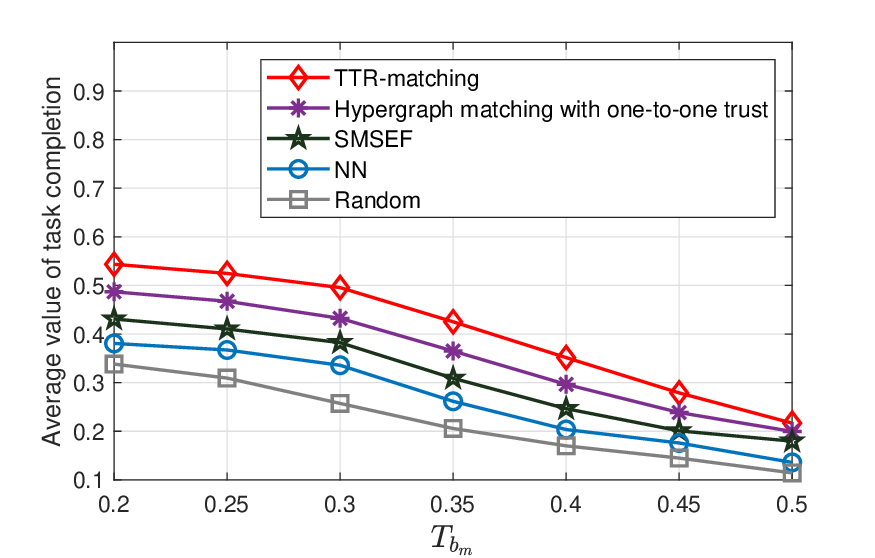}} \hspace{-0.29 in}
      \subfigure[$\xi_1 = 0.5$, $\xi_2 = 0.5$]{\includegraphics[scale=0.43]{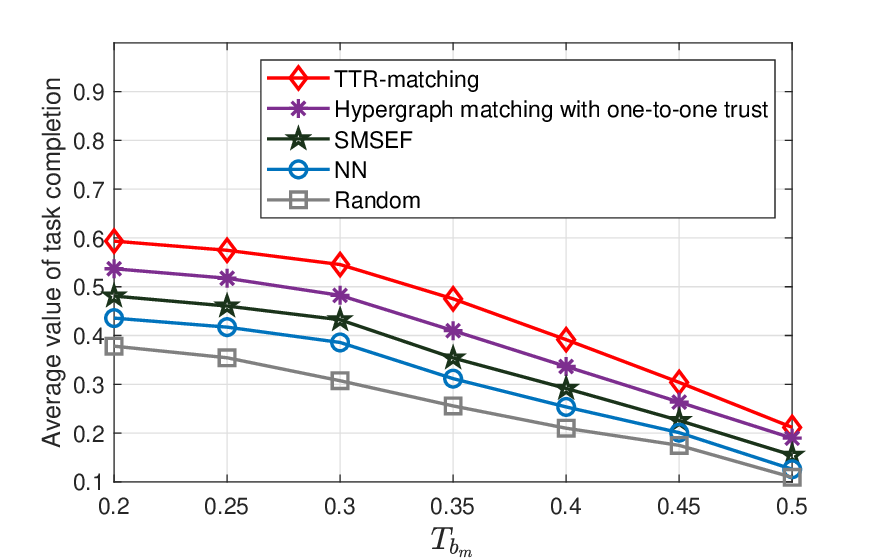}}
      \hspace{-0.29 in}
      \subfigure[$\xi_1 = 1$, $\xi_2 = 0$]{\includegraphics[scale=0.43]{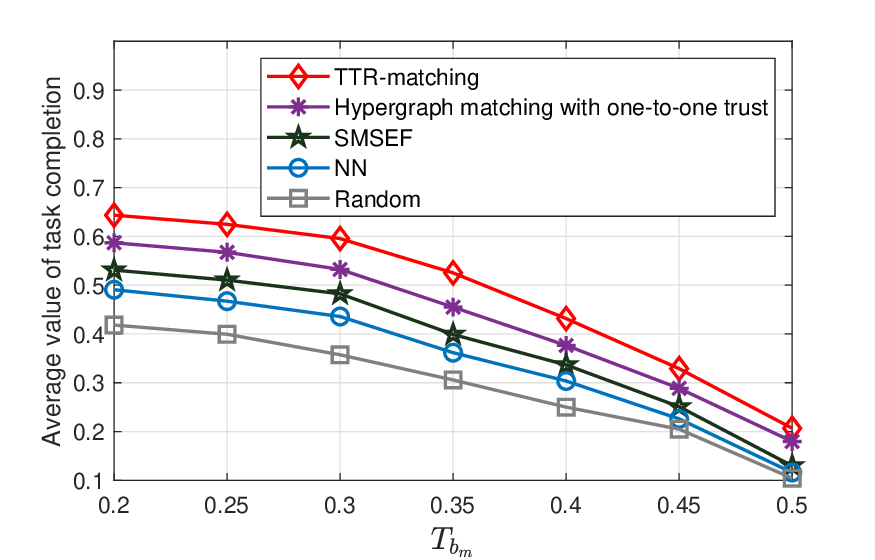}}
     \caption{Comparison of the average value of task completion when changing $T_{b_m}$.}
     \label{trust_demand}
\end{figure*}

\begin{table}[t] 
        \footnotesize  
	\centering
        \renewcommand{\arraystretch}{1.3}
	\caption{The features of $\bm{B}$}
	\label{task}
	  \begin{tabular}{p{0.8cm}<{\centering}|p{0.7cm}<{\centering}|p{0.9cm}<{\centering}|p{0.4cm}<{\centering}|p{0.7cm}<{\centering}|p{1cm}<{\centering}}
		\hline {Subtask} & $s$ & $d_{b_m}$ & $T_{b_m}$ & $t^{\text{max}}_{b_m}$ & $r_{b_m}$\\
		\hline\hline
             $b_1$ & 3DM &  5 MB & 0.2 & 0.6 s & 10 MB/s\\
             \hline
             $b_2$ & TWC & 1 MB & 0.2 & 0.6 s & 2 MB/s \\ 
             \hline
             $b_3$ & FR & 5 MB & 0.2 & 0.6 s & 10 MB/s\\
             \hline
   \end{tabular}
\end{table}

Additionally, it is observed that the difference between task-specific trust values and one-to-one trust values in Fig.~\ref{trust_comparison_0.1_0.1_0.8} is larger than that in Fig.~\ref{trust_comparison_0.1_0.8_0.1} and Fig.~\ref{trust_comparison_0.8_0.1_0.1}. This is because $T_{a_i,s,a_j}$ is assigned a larger weight in the trust calculation in Fig.~\ref{trust_comparison_0.1_0.1_0.8}, which is determined by different types of tasks. Conversely, since $T_{a_i,s,a_j}$ is assigned a smaller weight in Fig.~\ref{trust_comparison_0.1_0.8_0.1} and Fig.~\ref{trust_comparison_0.8_0.1_0.1}, the trust values obtained show only a slight difference compared to those obtained using the one-to-one trust model. Therefore, to more effectively highlight the advantages of the proposed model, assigning a larger weight to $T_{a_i,s,a_j}$ is preferable.


\vspace{-0.07 in}
\subsection{Comparison of the Average Value of Task Completion When Changing $T_{b_m}$}

This subsection investigates the impact of changing $T_{b_m}$ on the average value of task completion. In the task-specific trust model, the values of $\beta_1$, $\beta_2$, and $\beta_3$ are set to 0.2, 0.1, and 0.7, respectively. For one-to-one trust, $\beta_1$ and $\beta_2$ are set to 0.2 and 0.8. $a_1$ is designated as the task initiator, and all other devices serve as collaborators. Nearest neighbor (NN) search, random search, and SMSEF~\cite{8543056} are employed as baseline methods for comparison.

In Fig.~\ref{trust_demand}~(a), the values of $\xi_{1}$ and $\xi_{2}$ are set to 0 and 1, respectively, indicating that the value of task completion is determined solely by the value of energy. When the minimum trust requirement $T_{b_m}$ is 0.2, most collaborators meet this threshold. Consequently, $a_1$ can identify suitable collaborators across all methods to maximize the value of task completion. The proposed TTR-matching approach achieves the highest value compared to the other methods. Additionally, the one-to-one trust method performs well, outperforming SMSEF, NN, and random approaches. As $T_{b_m}$ increases, the number of collaborators meeting the trust threshold gradually decreases. This means that all methods may struggle to select the most suitable collaborators, resulting in a reduction in the value of task completion. However, the proposed TTR-matching method consistently achieves the highest value of task completion. In Fig.~\ref{trust_demand}~(c), $\xi_{1}$ and $\xi_{2}$ are assigned values of 1 and 0, respectively, signifying that the value of task completion is exclusively determined by the value of time. We can observe the same phenomenon as in Fig.~\ref{trust_demand}~(a), where the value of task completion decreases as $T_{b_m}$ increases for all algorithms. However, the proposed TTR-matching method consistently outperforms the comparison methods.


\subsection{Comparison of the Average Value of Task Completion When Changing $r_{b_m}$}
\begin{figure}[!t]
\centering
\includegraphics[scale=0.6]{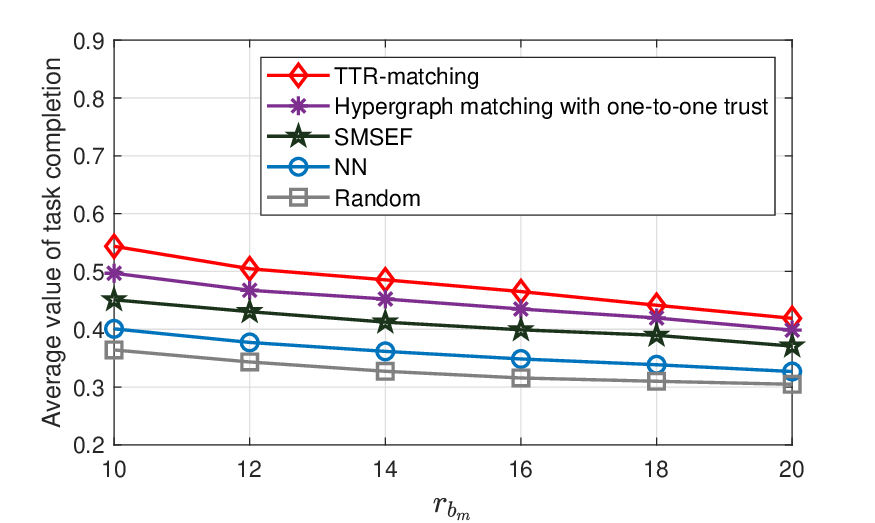}
\caption{Comparison of the average value of task completion when changing $r_{b_m}$.}
\label{transmission}
\end{figure}

This subsection investigates the impact of the minimum transmission rate on the value of task completion. Both $\xi_1$ and $\xi_2$ are set to 0.5. As shown in Fig.~\ref{transmission}, it can be observed that when $r_{b_m} = 10$ MB/s, each method achieves a relatively high value of task completion. This is because almost all collaborators meet the minimum transmission rate requirement, allowing $a_1$ to select the most suitable collaborators to perform subtasks. As $r_{b_m}$ increases, the number of eligible collaborators decreases, which reduces the selection pool for $a_1$ and leads to a decrease in the value of task completion. 
However, the proposed TTR-matching method consistently outperforms the comparison algorithms in achieving the highest value of task completion.

These experimental results demonstrate that the proposed TTR-matching method effectively promotes device collaboration through the task-specific trust model and precise task-resource matching.

\vspace{-0.05 in}
\subsection{Performance Evaluation of TTR-Matching in the Large-Scale System}
\begin{figure}[!t]
\centering
\includegraphics[scale=0.54]{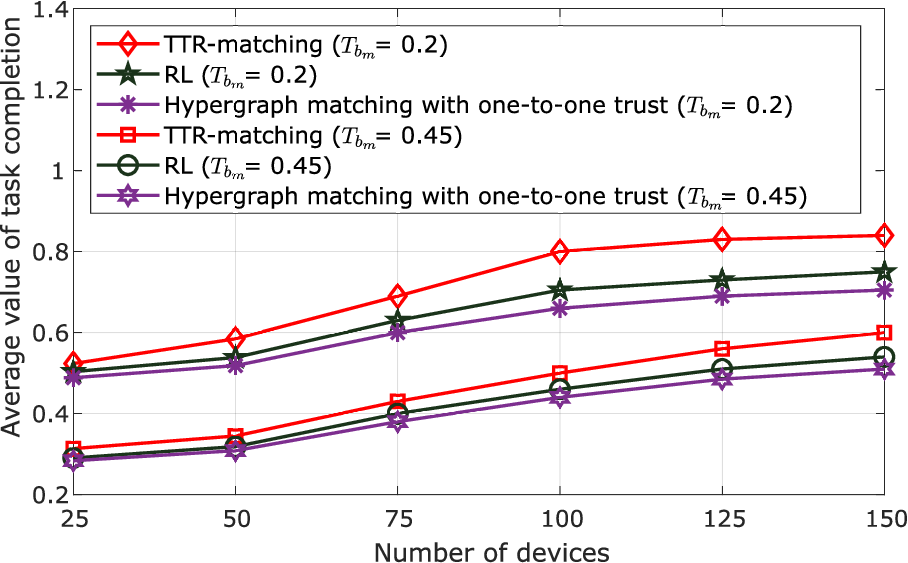}
\caption{Comparison of the average value of task completion when changing the number of devices.}
\label{number_devices}
\end{figure}

Device execution data from previous experiments involving Google Pixel 8, DELL 5200, DELL 5820, Lambda, and Rosbot are used to build device models for each device type. These device models are deployed in equal numbers on the NS3 simulation platform, serving as collaborators.
Device $a_1$ acts as the task initiator, generating tasks composed of four heterogeneous subtasks. The parameters of subtasks $b_1$, $b_2$, and $b_3$ are specified in Table~\ref{task}, while subtask $b_4$ is configured as follows: task type $s =$ VT, data size $d_{b_4} = 10 $ MB, maximum task completion tolerance time $t^{\text{max}}_{b_4} = 0.6$ s, and minimum transmission rate demand $r_{b_4} = 10$ MB/s. By varying the number of device models, we evaluate the proposed TTR-matching approach against a reinforcement learning-based method (DQN) ~\cite{9467050} and the one-to-one trust model. As shown in Fig.~\ref{number_devices}, the x-axis denotes the total number of devices, while the y-axis represents the average value of task completion. It can be observed that the average value of task completion increases with the number of devices across all methods. This is mainly due to the larger number of available devices for selection. Under the minimum trust demand of $T_{b_m} = 0.2$, the growth in the value of task completion tends to slow down once the number of devices exceeds 100. This slowdown occurs because additional devices contribute less significantly to improving the average value of task completion. In addition, the values of task completion at $T_{b_m} = 0.2$ are clearly higher than those at $T_{b_m} = 0.45$. This is because a lower trust demand enables more devices to meet the selection criteria. Our method clearly outperforms the baseline algorithms in achieving higher value of task completion. This demonstrates its effectiveness even in large-scale systems.

\begin{figure}[!t]
\centering
\includegraphics[scale=0.4]{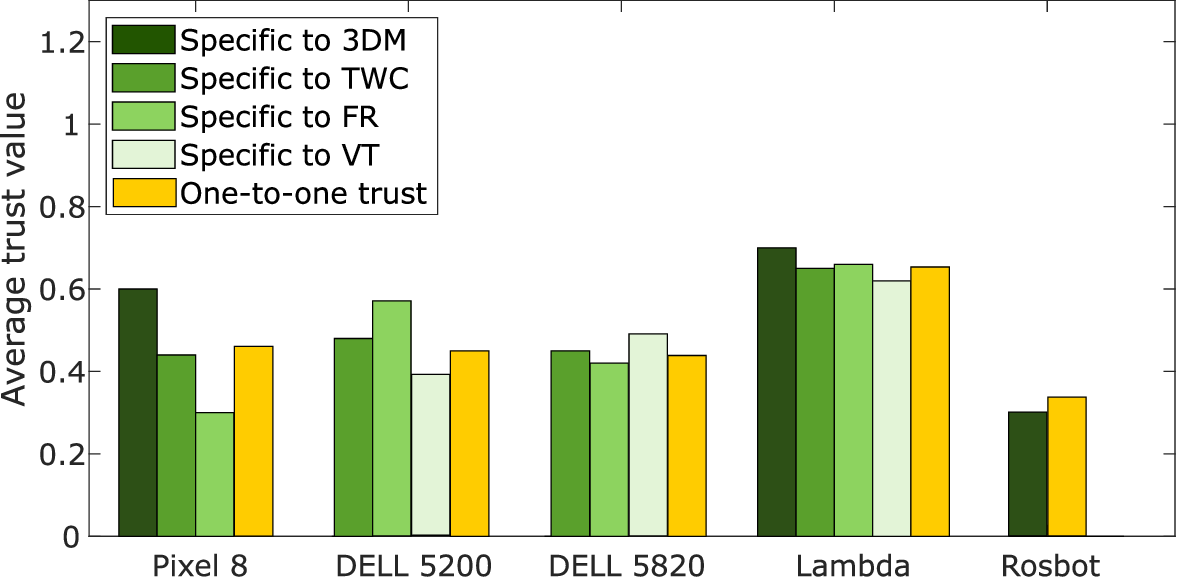}
\caption{Comparison of average trust values of devices.}
\label{trust_distribution}
\end{figure}

After completing 4,000 tasks, we calculate the average task-specific trust values for each device type and compare them with the one-to-one trust. All trust evaluations are conducted from the perspective of device $a_1$. In Fig.~\ref{trust_distribution}, the results show that the proposed approach can effectively capture trust differences across task types, revealing greater trust diversity compared to one-to-one trust. This finer-grained task-specific trust provides stronger differentiation, which helps in selecting more suitable and trustworthy devices. As a result, it contributes to improved value of task completion, as confirmed by the earlier experimental results. The results in Fig.~\ref{trust_distribution} further show that devices with different physical attributes are good at different types of tasks, which emphasizes the necessity of effectively matching tasks with appropriate devices.

\section{Conclusion}
\label{conclusion}
This paper has presented an innovative networked physical computing system that incorporates the physical attributes of resources and computing tasks, as well as task-specific trust relationships among devices to achieve value-oriented task completion. To implement this system, we have proposed a TTR-matching model in this research. First, a task-specific trusted physical resource hypergraph is designed, seamlessly integrating task-specific trust, resources, and tasks while representing their high-order collaborative relationships. In addition, a task hypergraph is developed to associate the task initiator with computing tasks. Furthermore, a hypergraph-aided matching method is utilized to align tasks and resources between the task hypergraph and the task-specific trusted physical resource hypergraph to maximize the value of task completion. Experimental findings indicate that the proposed TTR-matching model effectively captures dynamic and diverse trust relationships among devices and achieves the maximum value of task completion compared to other methods.

\section*{APPENDIX A \\ Experimental Setup}
\label{Appendixa}
Table~\ref{device} provides a detailed summary of the device features involved in the ICS. All devices are interconnected via WiFi, and their transmission power is calculated based on Google's WiFi energy consumption model~\cite{google}. These devices are equipped with various software to support diverse task types. For example, a device capable of supporting the task type of face recognition can receive images from the task initiator and execute a face detection program that counts the number of faces in each image. It is assumed that each task consists of several subtasks of different types, such as text, image, or video. The values of $\delta_1$, $\delta_2$, and $\delta_3$ are set to $1/3$, and the values of $\beta_1$, $\beta_2$, and $\beta_3$ are also set to $1/3$. $\eta^{\text{thr}}$ is set to $5\%$. Any pair of devices is connected via WiFi, and all parameters are obtained through the real-world measurement. The transmission rates and packet loss rates between all devices are measured using Iperf3~\cite{Iperf}. 
\begin{table}[!h] 
        \footnotesize  
	\centering
        \renewcommand{\arraystretch}{1.3}
	\caption{The features of devices}
	\label{device}
	  \begin{tabular}{p{1.9cm}<{\centering}|p{1.2cm}<{\centering}|p{1.2cm}<{\centering}|p{2.5cm}<{\centering}}
		\hline {Device} & {CPU} & $p_{a_i}$ (mW) & {Supported task types} \\
		\hline\hline
             iPad & 2.34 GHz & 700 & FR, TWC, 3DM, VT\\
             \hline
		     Google Pixel 8 & 2.91 GHz  & 740 & FR, TWC, 3DM\\
		\hline
              DELL 5200 & 3.8 GHz & 660 & FR, TWC, VT\\
            \hline
              DELL 5820 & 3.2 GHz & 660 & FR, TWC, VT\\ 
              \hline
              Lambda & 4.5 GHz & 660 & FR, TWC, VT, 3DM\\
              \hline
              Rosbot plus & 168 MHz  & 660 & 3DM\\
              \hline
              Robofleet & 72 MHz & 660 &  3DM\\
              \hline
   \end{tabular}
\end{table}


\footnotesize

\begin{thebibliography}{57}
\providecommand{\natexlab}[1]{#1}
\providecommand{\url}[1]{#1}
\csname url@samestyle\endcsname
\providecommand{\newblock}{\relax}
\providecommand{\bibinfo}[2]{#2}
\providecommand{\BIBentrySTDinterwordspacing}{\spaceskip=0pt\relax}
\providecommand{\BIBentryALTinterwordstretchfactor}{4}
\providecommand{\BIBentryALTinterwordspacing}{\spaceskip=\fontdimen2\font plus
\BIBentryALTinterwordstretchfactor\fontdimen3\font minus \fontdimen4\font\relax}
\providecommand{\BIBforeignlanguage}[2]{{%
\expandafter\ifx\csname l@#1\endcsname\relax
\typeout{** WARNING: IEEEtranN.bst: No hyphenation pattern has been}%
\typeout{** loaded for the language `#1'. Using the pattern for}%
\typeout{** the default language instead.}%
\else
\language=\csname l@#1\endcsname
\fi
#2}}
\providecommand{\BIBdecl}{\relax}
\BIBdecl

\bibitem[Fadlullah et~al.(2022)Fadlullah, Mao, and Kato]{9831429}
Z.~M. Fadlullah, B.~Mao, and N.~Kato, ``Balancing {QoS} and security in the edge: Existing practices, challenges, and {6G} opportunities with machine learning,'' \emph{{IEEE} Commun. Surveys Tuts.}, vol.~24, no.~4, pp. 2419--2448, Fourthquarter 2022.

\bibitem[Lv et~al.(2023)Lv, Xu, Nie, Yuan, Cai, Chen, and Xu]{9906430}
P.~Lv, W.~Xu, J.~Nie, Y.~Yuan, C.~Cai, Z.~Chen, and J.~Xu, ``Edge computing task offloading for environmental perception of autonomous vehicles in {6G} networks,'' \emph{{IEEE} Trans. Netw. Sci. Eng.}, vol.~10, no.~3, pp. 1228--1245, May-Jun. 2023.

\bibitem[Sharma and Wang(2019)]{8632974}
S.~K. Sharma and X.~Wang, ``Collaborative distributed {Q}-learning for {RACH} congestion minimization in cellular {IoT} networks,'' \emph{{IEEE} Commun. Lett.}, vol.~23, no.~4, pp. 600--603, Apr. 2019.

\bibitem[Ni et~al.(2024)Ni, Ao, Tian, Eldar, and Niyato]{10453339}
W.~Ni, H.~Ao, H.~Tian, Y.~C. Eldar, and D.~Niyato, ``Fedsl: Federated split learning for collaborative healthcare analytics on resource-constrained wearable {IoMT} devices,'' \emph{{IEEE} Internet Things J.}, vol.~11, no.~10, pp. 18\,934--18\,935, May 2024.

\bibitem[Xu et~al.(2023{\natexlab{a}})Xu, Niyato, Chen, Zhang, Kang, Xiong, Mao, and Han]{10177684}
M.~Xu, D.~Niyato, J.~Chen, H.~Zhang, J.~Kang, Z.~Xiong, S.~Mao, and Z.~Han, ``Generative {AI}-empowered simulation for autonomous driving in vehicular mixed reality metaverses,'' \emph{{IEEE} J. Sel. Topics Signal Process.}, vol.~17, no.~5, pp. 1064--1079, Sept. 2023.

\bibitem[Liu et~al.(2023{\natexlab{a}})Liu, Du, Niyato, Feng, Kang, and Xiong]{10144502}
Y.-J. Liu, H.~Du, D.~Niyato, G.~Feng, J.~Kang, and Z.~Xiong, ``Slicing4{M}eta: An intelligent integration architecture with multi-dimensional network resources for metaverse-as-a-service in web 3.0,'' \emph{{IEEE} Commun. Mag.}, vol.~61, no.~8, pp. 20--26, Aug. 2023.

\bibitem[Zhu and Wang(2025{\natexlab{a}})]{icc_botao_2}
B.~Zhu and X.~Wang, ``Accurate trust evaluation for effective operation of social {IoT} systems via hypergraph-enabled self-supervised contrastive learning,'' in \emph{Proc. IEEE Int. Conf. Commun. (ICC)}, 2025, pp. 1--6.

\bibitem[Chen et~al.(2024)Chen, Wang, and Shen]{10320384}
J.~Chen, X.~Wang, and X.~Shen, ``{RTE}: Rapid and reliable trust evaluation for collaborator selection and time-sensitive task handling in {I}nternet of {V}ehicles,'' \emph{{IEEE} Internet Things J.}, vol.~11, no.~7, pp. 12\,278--12\,291, Apr. 2024.

\bibitem[Xu et~al.(2023{\natexlab{b}})Xu, Zhang, Li, Yu, Leung, and Ji]{10104094}
Y.~Xu, H.~Zhang, X.~Li, F.~R. Yu, V.~C. Leung, and H.~Ji, ``Trusted collaboration for {MEC}-enabled {VR} video streaming: A multi-agent reinforcement learning approach,'' \emph{{IEEE} Trans. Veh. Technol.}, vol.~72, no.~9, pp. 12\,167--12\,180, Sept. 2023.

\bibitem[Zhu et~al.(2025)Zhu, Wang, Zhang, and Shen]{chain_of_trsut}
B.~Zhu, X.~Wang, L.~Zhang, and X.~S. Shen, ``Chain-of-trust: A progressive trust evaluation framework enabled by {G}enerative {AI},'' \emph{{IEEE} Netw.}, Jun. 2025, {E}arly {A}ccess, doi: 10.1109/MNET.2025.3582407.

\bibitem[Yang et~al.(2019)Yang, Hu, Chen, and Cao]{8516361}
Y.~Yang, W.~Hu, X.~Chen, and G.~Cao, ``Energy-aware {CPU} frequency scaling for mobile video streaming,'' \emph{{IEEE} Trans. Mobile Comput.}, vol.~18, no.~11, pp. 2536--2548, Nov. 2019.

\bibitem[Asheralieva and Niyato(2021)]{8933071}
A.~Asheralieva and D.~Niyato, ``Learning-based mobile edge computing resource management to support public blockchain networks,'' \emph{{IEEE} Trans. Mobile Comput.}, vol.~20, no.~3, pp. 1092--1109, Mar. 2021.

\bibitem[Liu et~al.(2020)Liu, Yin, Zhang, Cao, and Cheng]{8340067}
L.~Liu, B.~Yin, S.~Zhang, X.~Cao, and Y.~Cheng, ``Deep learning meets wireless network optimization: Identify critical links,'' \emph{{IEEE} Trans. Netw. Sci. Eng.}, vol.~7, no.~1, pp. 167--180, Jan.-Mar. 2020.

\bibitem[Du et~al.(2023)Du, Wang, Niyato, Kang, Xiong, Shen, and Kim]{9999298}
H.~Du, J.~Wang, D.~Niyato, J.~Kang, Z.~Xiong, X.~Shen, and D.~I. Kim, ``Exploring attention-aware network resource allocation for customized metaverse services,'' \emph{{IEEE} Netw.}, vol.~37, no.~6, pp. 166--175, Nov. 2023.

\bibitem[Hui et~al.(2020)Hui, Su, Luan, Li, Mao, and Wu]{9275320}
Y.~Hui, Z.~Su, T.~H. Luan, C.~Li, G.~Mao, and W.~Wu, ``A game theoretic scheme for collaborative vehicular task offloading in {5G} {H}et{N}ets,'' \emph{{IEEE} Trans. Veh. Technol.}, vol.~69, no.~12, pp. 16\,044--16\,056, Dec. 2020.

\bibitem[Zhang et~al.(2024{\natexlab{a}})Zhang, Wu, Chen, and Chen]{10623874}
T.~Zhang, F.~Wu, Z.~Chen, and S.~Chen, ``Optimization of edge–cloud collaborative computing resource management for {I}nternet of {V}ehicles based on multiagent deep reinforcement learning,'' \emph{{IEEE} Internet Things J.}, vol.~11, no.~22, pp. 36\,114--36\,126, Nov. 2024.

\bibitem[Huang et~al.(2021)Huang, Zeng, Ota, Dong, Wang, and Xiong]{9262001}
S.~Huang, Z.~Zeng, K.~Ota, M.~Dong, T.~Wang, and N.~N. Xiong, ``An intelligent collaboration trust interconnections system for mobile information control in ubiquitous {5G} networks,'' \emph{{IEEE} Trans. Netw. Sci. Eng.}, vol.~8, no.~1, pp. 347--365, Jan.-Mar. 2021.

\bibitem[Zhu and Wang(2025{\natexlab{b}})]{icc_botao}
B.~Zhu and X.~Wang, ``Rapid and continuous trust evaluation for effective task collaboration through {S}iamese model,'' in \emph{Proc. IEEE Int. Conf. Commun. (ICC)}, 2025, pp. 1--6.

\bibitem[Zhang et~al.(2023)Zhang, Han, Liu, Martínez-García, and Peng]{9760005}
F.~Zhang, G.~Han, L.~Liu, M.~Martínez-García, and Y.~Peng, ``Deep reinforcement learning based cooperative partial task offloading and resource allocation for {IIoT} applications,'' \emph{{IEEE} Trans. Netw. Sci. Eng.}, vol.~10, no.~5, pp. 2991--3006, Sept.-Oct. 2023.

\bibitem[Zhu and Wang(2024)]{10546264}
B.~Zhu and X.~Wang, ``Hypergraph-aided task-resource matching for maximizing value of task completion in collaborative {IoT} systems,'' \emph{{IEEE} Trans. Mobile Comput.}, vol.~23, no.~12, pp. 12\,247--12\,261, Dec. 2024.

\bibitem[Chen and Wang(2022)]{9749874}
R.~Chen and X.~Wang, ``Situation-aware orchestration of resource allocation and task scheduling for collaborative rendering in {IoT} visualization,'' \emph{{IEEE} Sustain. Comput.}, vol.~7, no.~4, pp. 935--949, Oct.-Dec. 2022.

\bibitem[Su et~al.(2024)Su, Zhang, Li, and Zhang]{10024361}
Q.~Su, Q.~Zhang, W.~Li, and X.~Zhang, ``Primal-dual-based computation offloading method for energy-aware cloud-edge collaboration,'' \emph{{IEEE} Trans. Mobile Comput.}, vol.~23, no.~2, pp. 1534--1549, Feb. 2024.

\bibitem[Zhang et~al.(2022)Zhang, Peng, Yan, and Sun]{9672105}
X.~Zhang, M.~Peng, S.~Yan, and Y.~Sun, ``Joint communication and computation resource allocation in fog-based vehicular networks,'' \emph{{IEEE} Internet Things J.}, vol.~9, no.~15, pp. 13\,195--13\,208, Aug. 2022.

\bibitem[Kang et~al.(2022)Kang, Li, Nie, Liu, Xu, Xiong, Niyato, and Yan]{9785702}
J.~Kang, X.~Li, J.~Nie, Y.~Liu, M.~Xu, Z.~Xiong, D.~Niyato, and Q.~Yan, ``Communication-efficient and cross-chain empowered federated learning for artificial intelligence of things,'' \emph{{IEEE} Trans. Netw. Sci. Eng.}, vol.~9, no.~5, pp. 2966--2977, Sept.-Oct. 2022.

\bibitem[Long et~al.(2023)Long, Zhang, Deng, Pei, Ouyang, and Xia]{9931975}
S.~Long, Y.~Zhang, Q.~Deng, T.~Pei, J.~Ouyang, and Z.~Xia, ``An efficient task offloading approach based on multi-objective evolutionary algorithm in cloud-edge collaborative environment,'' \emph{{IEEE} Trans. Netw. Sci. Eng.}, vol.~10, no.~2, pp. 645--657, Mar.-Apr. 2023.

\bibitem[Hou and Zhao(2023)]{9783171}
C.~Hou and Q.~Zhao, ``Optimal task-offloading control for edge computing system with tasks offloaded and computed in sequence,'' \emph{{IEEE} Trans. Autom. Sci. Eng.}, vol.~20, no.~2, pp. 1378--1392, Apr. 2023.

\bibitem[Huang et~al.(2023)Huang, Ji, Zhang, and Li]{9967961}
X.~Huang, G.~Ji, B.~Zhang, and C.~Li, ``Platform profit maximization in {D2D} collaboration based multi-access edge computing,'' \emph{{IEEE} Trans. Wireless Commun.}, vol.~22, no.~7, pp. 4282--4295, Jul. 2023.

\bibitem[Wu et~al.(2023)Wu, Liu, Li, Tang, and Wang]{9810544}
D.~Wu, T.~Liu, Z.~Li, T.~Tang, and R.~Wang, ``Delay-aware edge-terminal collaboration in green {I}nternet of {V}ehicles: A multiagent soft actor-critic approach,'' \emph{IEEE Trans. Green Commun. Netw.}, vol.~7, no.~2, pp. 1090--1102, Jun. 2023.

\bibitem[Yu et~al.(2024)Yu, Lu, and Fu]{10422726}
Y.~Yu, Q.~Lu, and Y.~Fu, ``Dynamic trust management for the edge devices in industrial internet,'' \emph{{IEEE} Internet Things J.}, vol.~11, no.~10, pp. 18\,410--18\,420, May 2024.

\bibitem[Pratap et~al.(2024)Pratap, Dass, and Misra]{10103199}
S.~Pratap, P.~Dass, and S.~Misra, ``{CoTEV}: Trustworthy and cooperative task execution in {I}nternet of {V}ehicles,'' \emph{{IEEE} Trans. Mobile Comput.}, vol.~23, no.~4, pp. 2915--2926, Apr. 2024.

\bibitem[Li et~al.(2024)Li, Qin, Liu, and Yu]{10251781}
J.~Li, Z.~Qin, W.~Liu, and X.~Yu, ``Energy-aware and trust-collaboration cross-domain resource allocation algorithm for edge-cloud workflows,'' \emph{{IEEE} Internet Things J.}, vol.~11, no.~4, pp. 7249--7264, Feb. 2024.

\bibitem[Liu et~al.(2023{\natexlab{b}})Liu, Zhang, Yan, Zhou, Tian, and Zhang]{9794601}
Y.~Liu, C.~Zhang, Y.~Yan, X.~Zhou, Z.~Tian, and J.~Zhang, ``A semi-centralized trust management model based on blockchain for data exchange in {IoT} system,'' \emph{{IEEE} Trans. Serv. Comput.}, vol.~16, no.~2, pp. 858--871, Mar.-Apr. 2023.

\bibitem[Marche and Nitti(2021)]{9305298}
C.~Marche and M.~Nitti, ``Trust-related attacks and their detection: A trust management model for the social {IoT},'' \emph{{IEEE} Trans. Netw. Service Manag.}, vol.~18, no.~3, pp. 3297--3308, Sep. 2021.

\bibitem[Fortino et~al.(2023)Fortino, Fotia, Messina, Rosaci, and Sarnè]{9908525}
G.~Fortino, L.~Fotia, F.~Messina, D.~Rosaci, and G.~M.~L. Sarnè, ``A social edge-based {IoT} framework using reputation-based clustering for enhancing competitiveness,'' \emph{{IEEE} Trans. Comput. Social Syst.}, vol.~10, no.~4, pp. 2051--2060, Aug. 2023.

\bibitem[Huang et~al.(2024)Huang, Li, Xiao, Long, and Liu]{10158489}
M.~Huang, Z.~Li, F.~Xiao, S.~Long, and A.~Liu, ``Trust mechanism-based multi-tier computing system for service-oriented edge-cloud networks,'' \emph{{IEEE} Trans. Depend. Sec. Comput.}, vol.~21, no.~4, pp. 1639--1651, Jul.-Aug. 2024.

\bibitem[Wang et~al.(2025{\natexlab{a}})Wang, Li, Liu, Qiu, and Luo]{10843332}
J.~Wang, Z.~Li, H.~Liu, T.~Qiu, and H.~Luo, ``A trust-based computation offloading framework in mobile cloud-edge computing networks,'' \emph{{IEEE} Trans. Mobile Comput.}, vol.~24, no.~6, pp. 5370--5385, Jun. 2025.

\bibitem[Wang et~al.(2025{\natexlab{b}})Wang, Chang, Hou, Zhang, Zhao, and Zhu]{10972056}
Q.~Wang, H.~Chang, P.~Hou, H.~Zhang, H.~Zhao, and H.~Zhu, ``Pilot assignment in {NOMA}-aided cell-free m{MIMO} systems: A hypergraph coloring scheme,'' \emph{{IEEE} Wireless Commun. Lett.}, pp. 1--5, Apr. 2025, {E}arly Access.

\bibitem[Li et~al.(2018)Li, Chen, and Guo]{8491251}
Z.~Li, S.~Chen, and C.~Guo, ``Location-aware hypergraph coloring based spectrum allocation for {D2D} communication,'' in \emph{Proc. 15th Int. Symp. Wireless Commun. Syst. (ISWCS)}, 2018, pp. 1--6.

\bibitem[Li and Zhu(2024)]{10444009}
J.~Li and X.~Zhu, ``Sum-rate optimization algorithms for {RIS} aided {D2D} multicast system based on hypergraph,'' \emph{{IEEE} Wireless Commun. Lett.}, vol.~13, no.~5, pp. 1330--1334, May 2024.

\bibitem[Latif et~al.(2022)Latif, Amer, and Kwasinski]{9990528}
O.~A. Latif, M.~Amer, and A.~Kwasinski, ``Hypergraph theory for network slicing,'' in \emph{Proc. Int. Conf. Electr. Comput. Technol. Appl. (ICECTA)}, 2022, pp. 283--286.

\bibitem[Zhang et~al.(2024{\natexlab{b}})Zhang, Yi, and Ma]{10594714}
S.~Zhang, N.~Yi, and Y.~Ma, ``A survey of computation offloading with task types,'' \emph{{IEEE} Trans. Intell. Transp. Syst.}, vol.~25, no.~8, pp. 8313--8333, Aug. 2024.

\bibitem[Wen et~al.(2012)Wen, Zhang, and Luo]{6195685}
Y.~Wen, W.~Zhang, and H.~Luo, ``Energy-optimal mobile application execution: Taming resource-poor mobile devices with cloud clones,'' in \emph{Proc. IEEE Int. Conf. Comput. Commun. (INFOCOM)}, 2012, pp. 2716--2720.

\bibitem[Mei et~al.(2017)Mei, Li, and Li]{7852434}
J.~Mei, K.~Li, and K.~Li, ``Customer-satisfaction-aware optimal multiserver configuration for profit maximization in cloud computing,'' \emph{{IEEE} Sustain. Comput.}, vol.~2, no.~1, pp. 17--29, Jan.-Mar. 2017.

\bibitem[Wang et~al.(2020)Wang, Qiu, Sangaiah, Xu, and Liu]{8727478}
T.~Wang, L.~Qiu, A.~K. Sangaiah, G.~Xu, and A.~Liu, ``Energy-efficient and trustworthy data collection protocol based on mobile fog computing in {I}nternet of {T}hings,'' \emph{{IEEE} Trans. Ind. Informat.}, vol.~16, no.~5, pp. 3531--3539, May 2020.

\bibitem[Yan et~al.(2018)Yan, Li, Li, and Cao]{7858754}
J.~Yan, C.~Li, Y.~Li, and G.~Cao, ``Adaptive discrete hypergraph matching,'' \emph{{IEEE} Trans. Cybern.}, vol.~48, no.~2, pp. 765--779, Feb. 2018.

\bibitem[Hou and Pelillo(2018)]{8545827}
J.~Hou and M.~Pelillo, ``A game-theoretic hyper-graph matching algorithm,'' in \emph{Proc. IEEE Int. Conf. Pattern Recognit. (ICPR)}, 2018, pp. 1012--1017.

\bibitem[Rota~Bulò and Pelillo(2013)]{6330964}
S.~Rota~Bulò and M.~Pelillo, ``A game-theoretic approach to hypergraph clustering,'' \emph{{IEEE} Trans. Pattern Anal. Mach. Intell.}, vol.~35, no.~6, pp. 1312--1327, Jun. 2013.

\bibitem[Duchenne et~al.(2011)Duchenne, Bach, Kweon, and Ponce]{5871640}
O.~Duchenne, F.~Bach, I.-S. Kweon, and J.~Ponce, ``A tensor-based algorithm for high-order graph matching,'' \emph{{IEEE} Trans. Pattern Anal. Mach. Intell.}, vol.~33, no.~12, pp. 2383--2395, Dec. 2011.

\bibitem[Kwak et~al.(2015)Kwak, Kim, Lee, and Chong]{7264984}
J.~Kwak, Y.~Kim, J.~Lee, and S.~Chong, ``{DREAM}: Dynamic resource and task allocation for energy minimization in mobile cloud systems,'' \emph{{IEEE} J. Sel. Areas Commun.}, vol.~33, no.~12, pp. 2510--2523, Dec. 2015.

\bibitem[Kwak et~al.(2016)Kwak, Choi, Chong, and Mohapatra]{7091048}
J.~Kwak, O.~Choi, S.~Chong, and P.~Mohapatra, ``Processor-network speed scaling for energy–delay tradeoff in smartphone applications,'' \emph{{IEEE/ACM} Trans. Netw.}, vol.~24, no.~3, pp. 1647--1660, Jun. 2016.

\bibitem[Miettinen and Nurminen(2010)]{10.5555/1863103.1863107}
A.~P. Miettinen and J.~K. Nurminen, ``Energy efficiency of mobile clients in cloud computing,'' in \emph{Proc. 2nd USENIX Conf. Hot Top. Cloud Comput.}, USA, 2010, pp. 1--7.

\bibitem[Wang et~al.(2021)Wang, Zhou, Li, Shi, Chen, and Hanzo]{9301339}
K.~Wang, Y.~Zhou, J.~Li, L.~Shi, W.~Chen, and L.~Hanzo, ``Energy-efficient task offloading in massive {MIMO}-aided multi-pair fog-computing networks,'' \emph{{IEEE} Trans. Commun.}, vol.~69, no.~4, pp. 2123--2137, Apr. 2021.

\bibitem[Kosta et~al.(2012)Kosta, Aucinas, Hui, Mortier, and Zhang]{6195845}
S.~Kosta, A.~Aucinas, P.~Hui, R.~Mortier, and X.~Zhang, ``Thinkair: Dynamic resource allocation and parallel execution in the cloud for mobile code offloading,'' in \emph{Proc. IEEE Int. Conf. Comput. Commun. (INFOCOM)}, 2012, pp. 945--953.

\bibitem[Wei et~al.(2018)Wei, Chen, and Zou]{8543056}
F.~Wei, S.~Chen, and W.~Zou, ``A greedy algorithm for task offloading in mobile edge computing system,'' \emph{China Commun.}, vol.~15, no.~11, pp. 149--157, Nov. 2018.

\bibitem[Liang et~al.(2022)Liang, Yu, Liu, Griffith, and Golmie]{9467050}
F.~Liang, W.~Yu, X.~Liu, D.~Griffith, and N.~Golmie, ``Toward deep {Q}-network-based resource allocation in industrial {I}nternet of {T}hings,'' \emph{{IEEE} Internet Things J.}, vol.~9, no.~12, pp. 9138--9150, Jun. 2022.

\bibitem[Google()]{google}
\BIBentryALTinterwordspacing
Google. [Online]. Available: \url{https://android.googlesource.com/platform/frameworks/base/+/master/core/res/res/xml/power_profile.xml.}
\BIBentrySTDinterwordspacing

\bibitem[Iperf()]{Iperf}
\BIBentryALTinterwordspacing
Iperf. [Online]. Available: \url{https://iperf.fr/iperf-download.php}
\BIBentrySTDinterwordspacing

\end{thebibliography}

\end{document}